\documentclass[runningheads]{llncs}
\usepackage{graphicx}
\usepackage{amsmath,amssymb} 
\usepackage{multirow}
\usepackage{diagbox}
\usepackage{subcaption}
\usepackage[export]{adjustbox}
\usepackage{xcolor}
\usepackage{stackengine}
\usepackage[ruled,vlined]{algorithm2e}
\usepackage{hyperref}
\hypersetup{
    colorlinks=true,
    linkcolor=cyan,
    filecolor=magenta,      
    urlcolor=blue,
    pdftitle={Overleaf Example},
    pdfpagemode=FullScreen,
    }
\begin{document}
\title{Towards performant and reliable undersampled MR reconstruction via diffusion model sampling}
%
%
	\author{Cheng Peng\inst{1} \and
	Pengfei Guo\inst{1} \and
	S. Kevin Zhou\inst{2,3} \and
	Vishal Patel\inst{1} \and
	Rama Chellappa\inst{1}}
	
	\institute{
	Johns Hopkins University, MD, USA \and 
	Medical Imaging, Robotics, and Analytic Computing Laboratory and Engineering (MIRACLE) Center, School of Biomedical Engineering \& Suzhou Institute for Advance Research, University of Science and Technology of China, Suzhou, China \and 
    Key Lab of Intelligent Information Processing of Chinese Academy of Sciences (CAS),
    Institute of Computing Technology, CAS, Beijing, China \\
    \email{\{cpeng26,pguo4,vpatel36,rchella4\}@jhu.edu, s.kevin.zhou@gmail.com}}
\maketitle              

\begin{abstract}
Magnetic Resonance (MR) image reconstruction from under-sampled acquisition promises faster scanning time. To this end, current State-of-The-Art (SoTA) approaches leverage deep neural networks and supervised training to learn a recovery model. While these approaches achieve impressive performances, the learned model can be fragile on unseen degradation, e.g. when given a different acceleration factor. These methods are also generally deterministic and provide a single solution to an ill-posed problem; as such, it can be difficult for practitioners to understand the reliability of the reconstruction. We introduce DiffuseRecon, a novel diffusion model-based MR reconstruction method. DiffuseRecon guides the generation process based on the observed signals and a pre-trained diffusion model, and does not require additional training on specific acceleration factors. DiffuseRecon is stochastic in nature and generates results from a distribution of fully-sampled MR images; as such, it allows us to explicitly visualize different potential reconstruction solutions. Lastly, DiffuseRecon proposes an accelerated, coarse-to-fine Monte-Carlo sampling scheme to approximate the most likely reconstruction candidate. The proposed DiffuseRecon achieves SoTA performances reconstructing from raw acquisition signals in fastMRI and SKM-TEA. Code will be open-sourced at \url{www.github.com/cpeng93/DiffuseRecon}.
\end{abstract}
\section{Introduction}

Magnetic Resonance Imaging (MRI) is a widely used medical imaging technique. It offers several advantages over other imaging modalities, such as providing high contrast on soft tissues and introducing no harmful radiation during acquisition. However, MRI is also limited by its long acquisition time due to the underlying imaging physics and machine quality. This leads to various issues ranging from patient discomfort to limited accessibility of the machines.

An approach to shorten MR scanning time is by under-sampling the signal in k-space during acquisition and recovering it by performing a post-process reconstruction algorithm. Recovering unseen signal is a challenging, ill-posed problem, and there has been a long history of research in addressing undersampled MR reconstruction. In general, this problem is formulated as:
\setlength{\belowdisplayskip}{1pt} \setlength{\belowdisplayshortskip}{1pt}
\setlength{\abovedisplayskip}{1pt} \setlength{\abovedisplayshortskip}{1pt}
\begin{equation}\label{eq:mr_recon}
    y_{\textrm{recon}} = \arg\min_{y}\lVert \mathcal{M}\mathcal{F}y-x_{\textrm{obs}}\rVert + \lambda R(y),~~ s.t.~~ x_{\textrm{obs}}=\mathcal{M}x_{\textrm{full}},
\end{equation}
where $x_{\{\textrm{full,obs}\}}$ denotes the fully-sampled and under-sampled k-space signal, $\mathcal{M}$ denotes the undersampling mask, and $\mathcal{F}$ denotes the Fourier operator. The goal is to find an MR image $y$ such that its k-space content $\mathcal{M}\mathcal{F}y$ agrees with $x_{\textrm{obs}}$; this is often known as the data consistency term. Furthermore, $y_{\textrm{recon}}$ should follow certain prior knowledge about MR images, as expressed by the regularization term $R(*)$. The design of $R$ is subject to many innovations. Lustig et al.~\cite{lustig2007sparse} first proposed to use Compressed Sensing motivated $\ell_{1}$-minimization algorithm for MRI reconstruction, assuming that the undersampled MR images have a sparse representation in a transform domain. Ravishankar et al.~\cite{DBLP:journals/tmi/RavishankarB11} applied a more adaptive sparse modelling through Dictionary Learning, where the transformation is optimized through a set of data, leading to improved sparsity encoding. As interests grow in this field, more MR data has become publicly available. Consequently, advances in Deep Learning (DL), specifically with supervised learning, have been widely applied in MR reconstruction. Generally, DL-based methods~\cite{Eo2018-xi,DBLP:conf/cvpr/ZhouZ20,DBLP:conf/miccai/GuoVWZJP21,DBLP:journals/tmi/SchlemperCHPR18,DBLP:conf/miccai/RonnebergerFB15,Akcakaya2019-lm,8756028,7493320,8233175,https://doi.org/10.1002/mrm.26977,8425639,DBLP:conf/cvpr/SriramZMZDS20,DBLP:conf/cvpr/GuoWZJP21} train Convolutional Neural Networks (CNNs) with paired data $\{y_{\textrm{und}},y_{\textrm{full}}\}=\{\mathcal{F}^{-1}x_{\textrm{obs}},\mathcal{F}^{-1}x_{\textrm{full}}\}$. Following the formulation of Eq.~(\ref{eq:mr_recon}), data consistency can be explicitly enforced within the CNN by replacing $\mathcal{M}\mathcal{F}y$ with $x_{\textrm{obs}}$~\cite{DBLP:journals/tmi/SchlemperCHPR18}. The resulting CNN serves as a parameterized $R(*,\theta)$, and regularizes test images based on learned $\theta$ from the training distribution. 

While supervised DL-based methods have led to impressive results, these methods generally train CNNs under specific degradation conditions; e.g., the under-sampling mask $\mathcal{M}$ that generates $y_\textrm{und}$ follows a particular acceleration factor. As a consequence, when the acceleration factor changes, the performances of various models often degrade significantly, making $R(*,\theta)$ less reliable in general. Furthermore, while the trained CNNs infer the most likely estimation of $y_\textrm{full}$ given $y_\textrm{und}$, they do not provide possible alternative solutions. Since Eq.~\ref{eq:mr_recon} is an under-constrained problem, $y_{\textrm{und}}$ can have many valid solutions. The ability to observe different solutions can help practitioners understand the potential variability in reconstructions and make more robust decisions. As such, finding a \underline{stochastic} $R$ that is \underline{generally applicable} across all $x_{\textrm{obs}}$ is of high interest.


We leverage the recent progress in a class of generative methods called \textbf{diffusion models}~\cite{DBLP:conf/nips/HoJA20,DBLP:conf/icml/NicholD21,DBLP:journals/corr/abs-2105-05233}, which use a CNN to perform progressive reverse-diffusion and maps a prior Gaussian distribution $\mathcal{N}$ to a learned image distribution, e.g. $p_{\theta}(y_\textrm{full})$. Based on a pre-trained $\theta$, we propose to guide the iterative reverse-diffusion by gradually introducing $x_{\textrm{obs}}$ to the intermediate results, as shown in Fig. 1. This allows us to generate reconstructions in the marginal distribution $p_{\theta}(y_{\textrm{full}}|x_{\textrm{obs}})$, where any sample $y_{\textrm{recon}}\sim p_{\theta}(y_{\textrm{full}}|x_{\textrm{obs}})$ agrees with the observed signal and lies on the MR image distribution $p_{\theta}(y_{\textrm{full}})$. 

We make three contributions. 
(i) Our proposed \textbf{DiffuseRecon} performs MR reconstruction by gradually guiding the reverse-diffusion process with observed k-space signal and is robust to changing sampling condition using a single model. 
(ii) We propose a coarse-to-fine sampling algorithm, which allows us to estimate the most likely reconstruction and its variability within $p_{\theta}(y_{\textrm{full}}|x_{\textrm{obs}})$ while leading to an approximately 40$\times$ speed-up compared to naive sampling.
(iii) We perform extensive experiments using raw acquisition signals from fastMRI~\cite{DBLP:journals/corr/abs-1811-08839} and SKM-TEA~\cite{desai2021skm}, and demonstrate superior performance over SoTA methods.
\begin{figure}[t]
    \centering
      \includegraphics[width=1\textwidth]{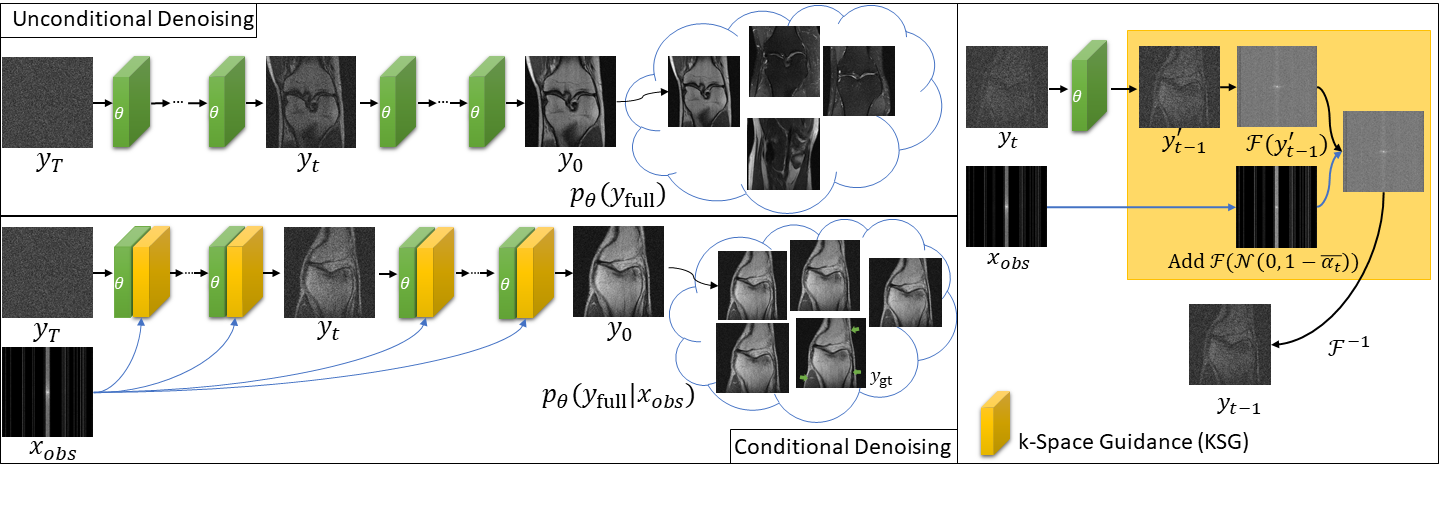}
\vspace{-3em}
    \caption{DiffuseRecon gradually incorporates $x_{\textrm{obs}}$ into the denoising process through a k-Space Guidance (KSG) module; as such, we can directly generate samples from $p_{\theta}(y_{\textrm{full}}|x_{\textrm{obs}})$. Visualizations are based on 8X undersampling.}
    \label{fig:network}
\vspace{-1em}
\end{figure}

\section{DiffuseRecon}

\textbf{Background.} Diffusion model~\cite{DBLP:conf/nips/HoJA20,DBLP:conf/icml/NicholD21,DBLP:journals/corr/abs-2105-05233} is a class of unconditional generative methods that aims to transform a Gaussian distribution to the empirical data distribution. Specifically, the forward diffusion process is a Markov Chain that gradually adds Gaussian noise to a clean image, which can be expressed as:
\begin{align} \label{eq:forward}
    q(y_t|y_{t-1})=\mathcal{N}(y_t;\sqrt{1-\beta_{t}}y_{t-1},\beta_{t}\mathbf{I}); q(y_t|y_0)=\mathcal{N}(y_t;\sqrt{\bar{\alpha}_t}y_0,(1-\bar{\alpha}_t)\mathbf{I}),
\end{align}
where $y_t$ denotes the intermediate noisy images, $\beta_{t}$ denotes the noise variance, $\alpha_t=1-\beta_t$, and $\bar{\alpha}_t=\prod_{s=1}^{t}\alpha_s$. For simplicity, $\beta_{t}$ follows a fixed schedule. When $T$ is sufficiently large, $y_T$ is dominated by noise. A CNN model is introduced to gradually reverse the forward diffusion process, i.e. \emph{denoise} $y_t$, by estimating
\begin{align} \label{eq:backward}
    p_\theta(y_{t-1}|y_{t})=\mathcal{N}(y_{t-1};\epsilon_{\theta}(y_t,t),\sigma_{t}^2\mathbf{I}),
\end{align}
where $\sigma_{t}^2=\bar{\beta_{t}}=\frac{1-\bar{\alpha}_{t-1}}{1-\bar{\alpha}_{t}}\beta_t$ in this work. With fixed variance, $\epsilon_{\theta}$ in Eq.~(\ref{eq:backward}) is trained by mean-matching the diffusion noise through an $\mathcal{L}_2$ loss. We follow~\cite{DBLP:conf/icml/NicholD21} with some modifications to train a diffusion model that generates MR images. 

After training on MR images $y_{\textrm{full}}$, $\epsilon_{\theta}$ can be used to gradually transform noise into images that follow the distribution $p_{\theta}(y_{\textrm{full}})$, as shown in Fig.~\ref{fig:network}. We note that diffusion models generate images \emph{unconditionally} by their original design. To reconstruct high fidelity MR images \emph{conditioned} on $x_{obs}$, DiffuseRecon is proposed to gradually modify $y_t$ such that $y_0$ agrees with $x_{obs}$. DiffuseRecon consists of two parts, k-Space Guidance and Coarse-to-Fine Sampling.

\noindent {\bf{K-Space Guidance. }} We note that $R(*)$ is naturally enforced by following the denoising process with a pre-trained $\theta$, as the generated images already follow the MR data distribution. As such, k-Space Guidance (KSG) is proposed to ensure the generated images follow the data consistency term. In a diffusion model, a denoised image is generated at every step $t$ by subtracting the estimated noise from the previous $y_t$, specifically:
\begin{align} \label{eq:denoise_ksg}
    y^{\prime}_{t-1} =\frac{1}{\sqrt{\alpha_t}}(y_t-\frac{1-\alpha_t}{\sqrt{1-\bar{\alpha}_t}}\epsilon_\theta(y_t,t))+\sigma_t\mathbf{z}, \mathbf{z}\sim\mathcal{N}(0,\mathbf{I}).
\end{align}
For unconditional generation, $y_{t-1}=y^{\prime}_{t-1}$, and the process repeats until $t=0$. For DiffuseRecon, KSG gradually mixes observed k-space signal $x_{\textrm{obs}}$ with $y^{\prime}_{t-1}$. To do so, KSG first adds a zero-mean noise on $x_{\textrm{obs}}$ to simulate its diffused condition at step $t$. The noisy observation is then mixed with $y^{\prime}_{t-1}$ in k-space based on the undersampling mask $\mathcal{M}$. This process is expressed as follows:
\begin{equation}
\begin{aligned} \label{eq:denoise}
    &y_{t} = \mathcal{F}^{-1}((1-\mathcal{M})\mathcal{F}y^{\prime}_{t}+\mathcal{M}x_{\textrm{obs},t}), \\
    &\text{where } x_{\textrm{obs},t} = x_{\textrm{obs}}+\mathcal{F}(\mathcal{N}(0,(1-\bar{\alpha}_t)\mathbf{I})).
\end{aligned}
\end{equation}

The resulting $y_{t-1}$ is iteratively denoised with decreasing $t$ until $y_0$ is obtained, and $x_{\textrm{obs},0} = x_{\textrm{obs}}$. As such, $y_0$ shares the same observed k-space signal and achieves the data consistency term $\mathcal{M}\mathcal{F}y_0=x_{\textrm{obs}}$. Since all generated samples fall on the data distribution $p_\theta(y_{\textrm{full}})$, KSG allows us to stochastically sample from the marginal distribution $p_\theta(y_{\textrm{full}}|x_{\textrm{obs}})$ through a pre-trained diffusion model. As demonstrated in Fig. 2a, we can observe the variations in $y_0$ with a given $x_{\textrm{obs}}$ and determine the reliability of the reconstructed results. Furthermore, KSG is parameter-free and applicable to \emph{any} undersampling mask $\mathcal{M}$ without finetuning.


\captionsetup[subfigure]{labelformat=empty}
\begin{figure*}[t]
    \setlength{\abovecaptionskip}{0pt}
    \setlength{\tabcolsep}{0.5pt}
    \centering
    \begin{tabular}[b]{cc}
        \begin{subfigure}[b]{0.65\linewidth}
            \hspace{-2pt}
            \begin{tabular}[b]{|ccccc|}
            \hline
            Mask & Init. 1 & Init. 2 & Init. 3 & Var.\\
            \hline
    
            \begin{subfigure}[b]{0.10\linewidth}
                \includegraphics[width=\textwidth,height=\textwidth]{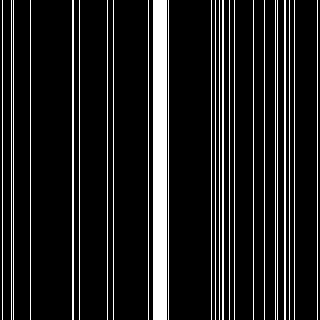}
            \end{subfigure} &
            \begin{subfigure}[b]{0.10\linewidth}
                \includegraphics[width=\textwidth,height=\textwidth]{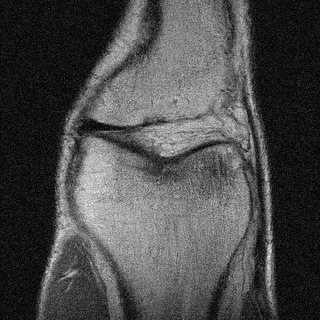}
            \end{subfigure} &
            \begin{subfigure}[b]{0.10\linewidth}
                \includegraphics[width=\textwidth,height=\textwidth]{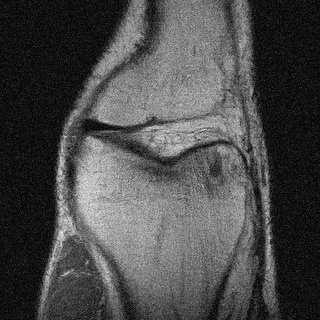}
            \end{subfigure} &
            \begin{subfigure}[b]{0.10\linewidth}
                \includegraphics[width=\textwidth,height=\textwidth]{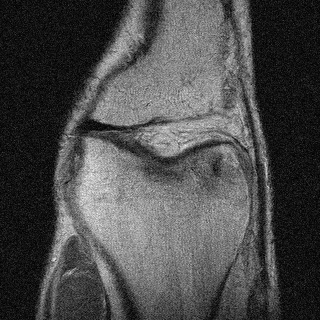}
            \end{subfigure} &
            \begin{subfigure}[b]{0.10\linewidth}
                \includegraphics[width=\textwidth,height=\textwidth]{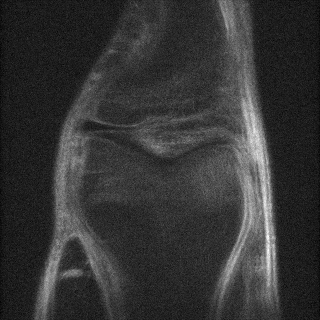}
            \end{subfigure}\\

            \hline        
            
            \begin{subfigure}[b]{0.10\linewidth}
                \includegraphics[width=\textwidth,height=\textwidth]{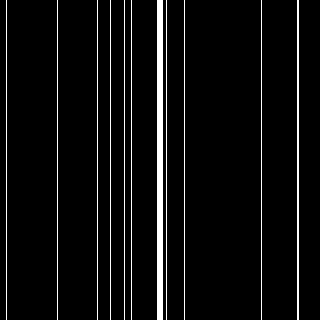}
            \end{subfigure} &
            \begin{subfigure}[b]{0.10\linewidth}
                \includegraphics[width=\textwidth,height=\textwidth]{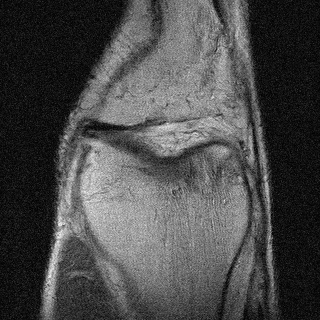}
            \end{subfigure} &
            \begin{subfigure}[b]{0.10\linewidth}
                \includegraphics[width=\textwidth,height=\textwidth]{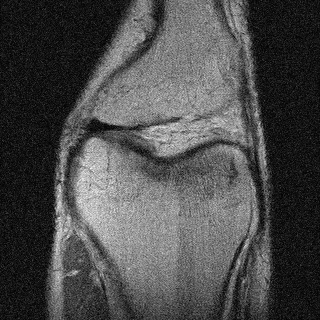}
            \end{subfigure} &
            \begin{subfigure}[b]{0.10\linewidth}
                \includegraphics[width=\textwidth,height=\textwidth]{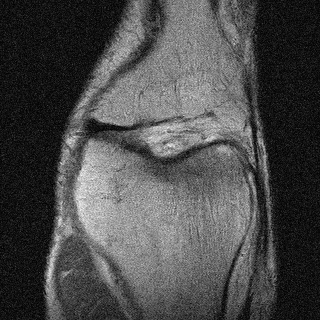}
            \end{subfigure} &
            \begin{subfigure}[b]{0.10\linewidth}
                \includegraphics[width=\textwidth,height=\textwidth]{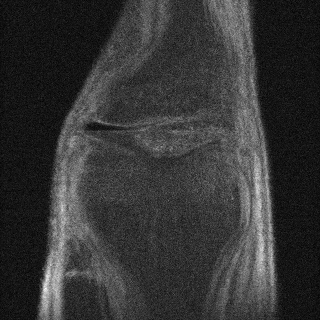}
            \end{subfigure}  \\
              
            \hline        
            
            \begin{subfigure}[b]{0.10\linewidth}
                \includegraphics[width=\textwidth,height=\textwidth]{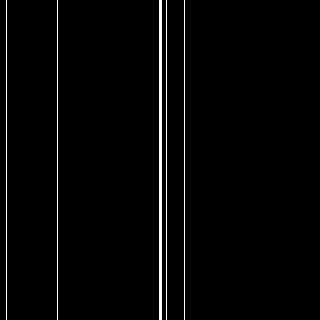}
            \end{subfigure} &
            \begin{subfigure}[b]{0.10\linewidth}
                \includegraphics[width=\textwidth,height=\textwidth]{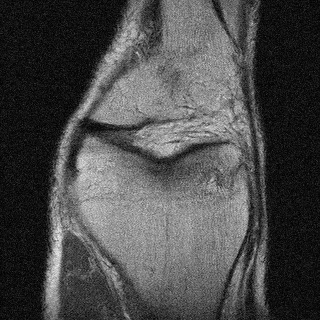}
            \end{subfigure} &
            \begin{subfigure}[b]{0.10\linewidth}
                \includegraphics[width=\textwidth,height=\textwidth]{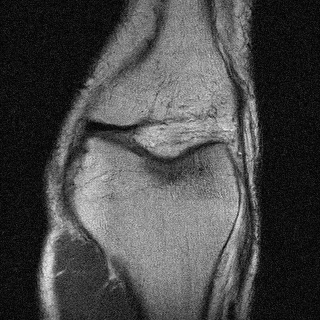}
            \end{subfigure} &
            \begin{subfigure}[b]{0.10\linewidth}
                \includegraphics[width=\textwidth,height=\textwidth]{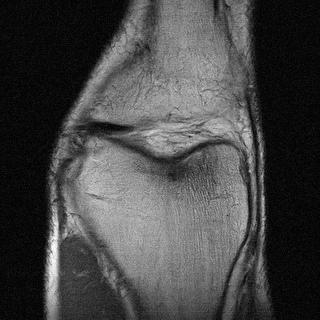}
            \end{subfigure} &
            \begin{subfigure}[b]{0.10\linewidth}
                \includegraphics[width=\textwidth,height=\textwidth]{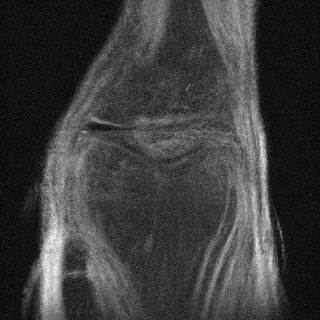}
            \end{subfigure}  \\
            
            \hline       
            \end{tabular}  
            \caption{\hspace{-10em}(a)}
        \end{subfigure}
        \hspace{-11.3em}
        \begin{subfigure}[b]{0.65\linewidth}    
            \includegraphics[width=\textwidth]{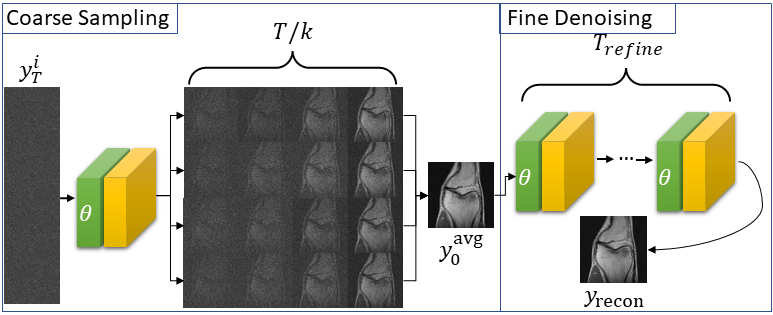}
            \caption{(b)}
        \end{subfigure}
        
    \end{tabular}     
    \caption{(a) Visualizations on various under-sampled k-space signal (8$\times$, 16$\times$, 32$\times$, 1D Gaussian Mask). (b) To accelerate the sampling process for MC simulation, multiple coarse samples are generated using $\frac{T}{k}$ steps. These samples are averaged to $y^{\textrm{avg}}_{0}$ and reintroduced to the denoising network for $T_{\textrm{refine}}$ steps.}
    \label{fig:C2F}
    \vspace{-2em}
\end{figure*}
\noindent {\bf{Coarse-to-Fine Sampling. }} While sampling in $p_\theta(y_{\textrm{full}}|x_{\textrm{obs}})$ allows us to generate different reconstruction candidates, selecting the best reconstruction out of all candidates is a challenge.
Since we do not have an analytic solution for $p_\theta(y_{\textrm{full}}|x_{\textrm{obs}})$, Monte-Carlo (MC) sampling can be used to estimate $\mathbb{E}(p_\theta(y_{\textrm{full}}|x_{\textrm{obs}}))$ in an unbiased way with sufficient samples. However, as sampling from a diffusion model already requires multiple steps, generating multiple samples is computationally costly. 
In practice, the denoising process can be accelerated by evenly spacing the $\{T,T-1,...,1\}$ schedule to a shorter schedule $\{T,T-k,...,1\}$~\cite{DBLP:conf/icml/NicholD21}, where $k>1$, and modifying the respective weights $\beta, \alpha$ based on $k$. However, the acceleration tends to produce less denoised results when $k$ is large.

We propose a Coarse-to-Fine (C2F) sampling algorithm to greatly accelerate the MC sampling process without loss in performance. Specifically, we note that the diffusion noise that is added in Eq.~(\ref{eq:forward}) is zero-mean with respect to $y_0$, and can be reduced by averaging multiple samples. Since multiple samples are already required to estimate $\mathbb{E}(p_\theta(y_{\textrm{full}}|x_{\textrm{obs}}))$, we leverage the multi-sample scenario to more aggressively reduce the denoising steps. As shown in Fig. 2b, C2F sampling creates $N$ instances of $y^{i,\frac{T}{k}}_T\sim\mathcal{N}(0,\mathbf{I})$ and applies denoising individually for $\frac{T}{k}$ steps based on a re-spaced schedule. The noisy results are averaged to produce $y^{\textrm{avg}}_0 = \frac{1}{N}\sum^{N}_{i=0}y^{i,\frac{T}{k}}_0$. Finally,  $y^{\textrm{avg}}_0$ is refined by going through additional $T_{\textrm{refine}}$ steps with $\epsilon_\theta$. To control the denoising strength in $\epsilon_\theta$, $\{T_{\textrm{refine}},T_{\textrm{refine}}-1,...,1\} \in \{T,T-1,...,1\},T_{\textrm{refine}}\ll T$. The last refinement steps help remove blurriness introduced by averaging multiple samples and lead to more realistic reconstruction results. During the refinement steps, $x_{\textrm{obs}}$ directly replaces k-space signals in the reconstructions, as is consistent with $\mathcal{M}\mathcal{F}y^{i,\frac{T}{k}}_0$.

 Compared to a naive approach which requires $TN$ total steps to estimate $\mathbb{E}(p_\theta(y_{\textrm{full}}|x_{\textrm{obs}}))$, C2F sampling introduces an approximately $k$-time speed-up. Furthermore, while $y^{i,\frac{T}{k}}_0$ are noisy compared to their fully-denoised version $y^{i,T}_0$, their noise is approximately Gaussian and introduces only a constant between $Var(y^{i,\frac{T}{k}}_0)$ and $Var(y^{i,T}_0)$. As such, given a reasonable $N$, variance in $p_\theta(y_{\textrm{full}}|x_{\textrm{obs}})$ can still be estimated well from coarse samples. 



\section{Experiments}

\textbf{Dataset.} In our experiments, we use two large knee MR datasets that contain raw, complex-value acquisition signals. FastMRI~\cite{DBLP:journals/corr/abs-1811-08839} contains 1172 subjects, with approximately 35 slices per subject and is split such that 973 subjects are used for training and 199 subjects are used for evaluation. SKM-TEA~\cite{desai2021skm} contains 155 subjects, with approximately 160 slices per subject; 134 subjects are used for training, 21 subjects are used for evaluation. Single-coil data is used for both datasets. Undersampling masks $\mathcal{M}$ are generated by using the masking function provided in the fastMRI challenge with 6$\times$ and 8$\times$ accelerations. To avoid slices with little information, the first five and ten slices are removed from evaluation for fastMRI and SKM-TEA respectively. To accommodate methods based on over-complete CNNs, which require upsampling and a larger memory footprint, images from SKM-TEA are center-cropped to $\mathbb{R}^{448\times448}$.

\hspace{-1.5em}\textbf{Implementation Details.} All baselines are implemented in PyTorch and trained with $\mathcal{L}_1$ loss and Adam optimizer. Following~\cite{DBLP:conf/miccai/GuoVWZJP21}, a learning rate is set to $1.5\times10^{-4}$ and reduced by $90\%$ every five epochs; to ensure a fair comparison, U-Net~\cite{DBLP:conf/miccai/RonnebergerFB15} is implemented with a data consistency layer at the end. For DiffuseRecon, we follow~\cite{DBLP:conf/icml/NicholD21} by using a cosine noise schedule and a U-Net architecture with multi-head attention as the denoising model. The model is modified to generate two consecutive slices; as such, the input and output channel sizes are 4, accounting for complex values. For C2F Sampling, $\{T,k,N,T_{\textrm{refine}}\}=\{4000,40,10,20\}$, which gives a $\frac{TN}{\frac{TN}{k}+T_{\textrm{refine}}}\approx39$ times speed-up compared to naive MC sampling. 


\captionsetup[subfigure]{labelformat=empty}
\begin{figure*}[htb!]
    \setlength{\abovecaptionskip}{0pt}
    \setlength{\tabcolsep}{0.5pt}
    \centering
    \begin{tabular}[b]{cc}

        \begin{subfigure}[b]{0.51\linewidth}    
            \includegraphics[width=\textwidth]{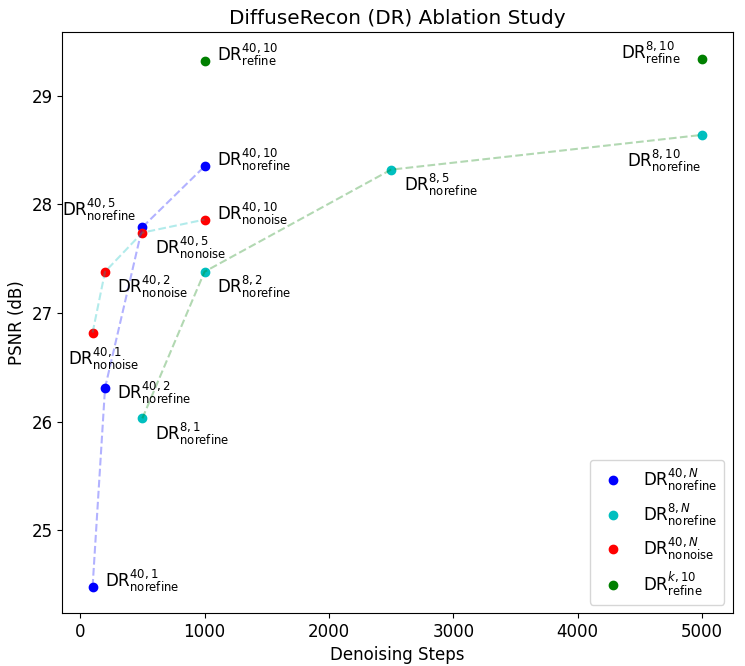}
            \caption{(a)}
        \end{subfigure}
        
        \begin{subfigure}[b]{0.73\linewidth}
        \begin{tabular}[b]{|c|c|c|c|}
            \hline
            D5C5 & OUCR & Ours & GT \\
            
            \hline
            \multicolumn{4}{|c|}{fastMRI, 8$\times$$\rightarrow$4$\times$}\\
        
            \begin{subfigure}[b]{0.160\linewidth}
                \includegraphics[width=\textwidth,height=\textwidth]{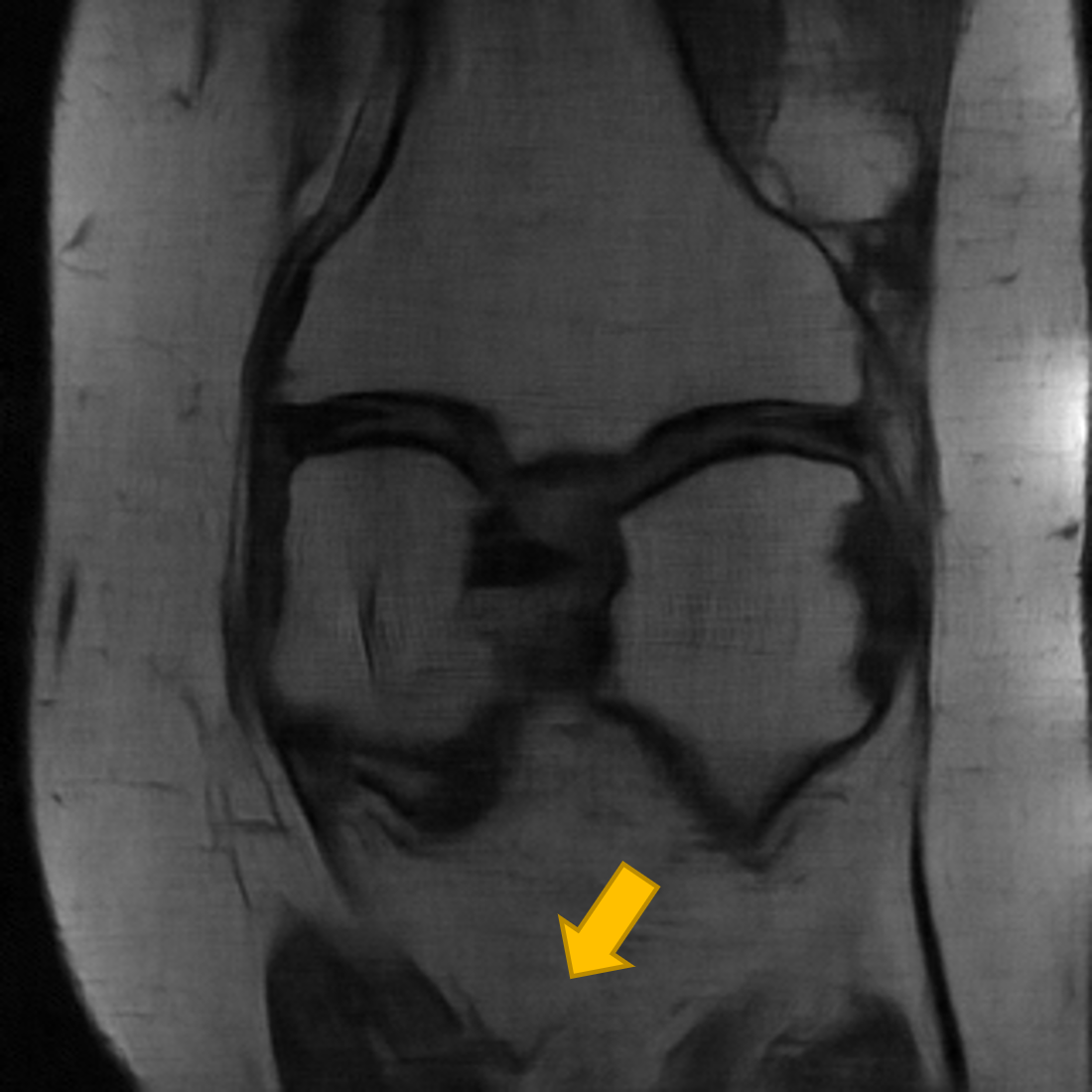}
                \caption{\scriptsize{33.1/.791}}
            \end{subfigure} &
            \begin{subfigure}[b]{0.160\linewidth}
                \includegraphics[width=\textwidth,height=\textwidth]{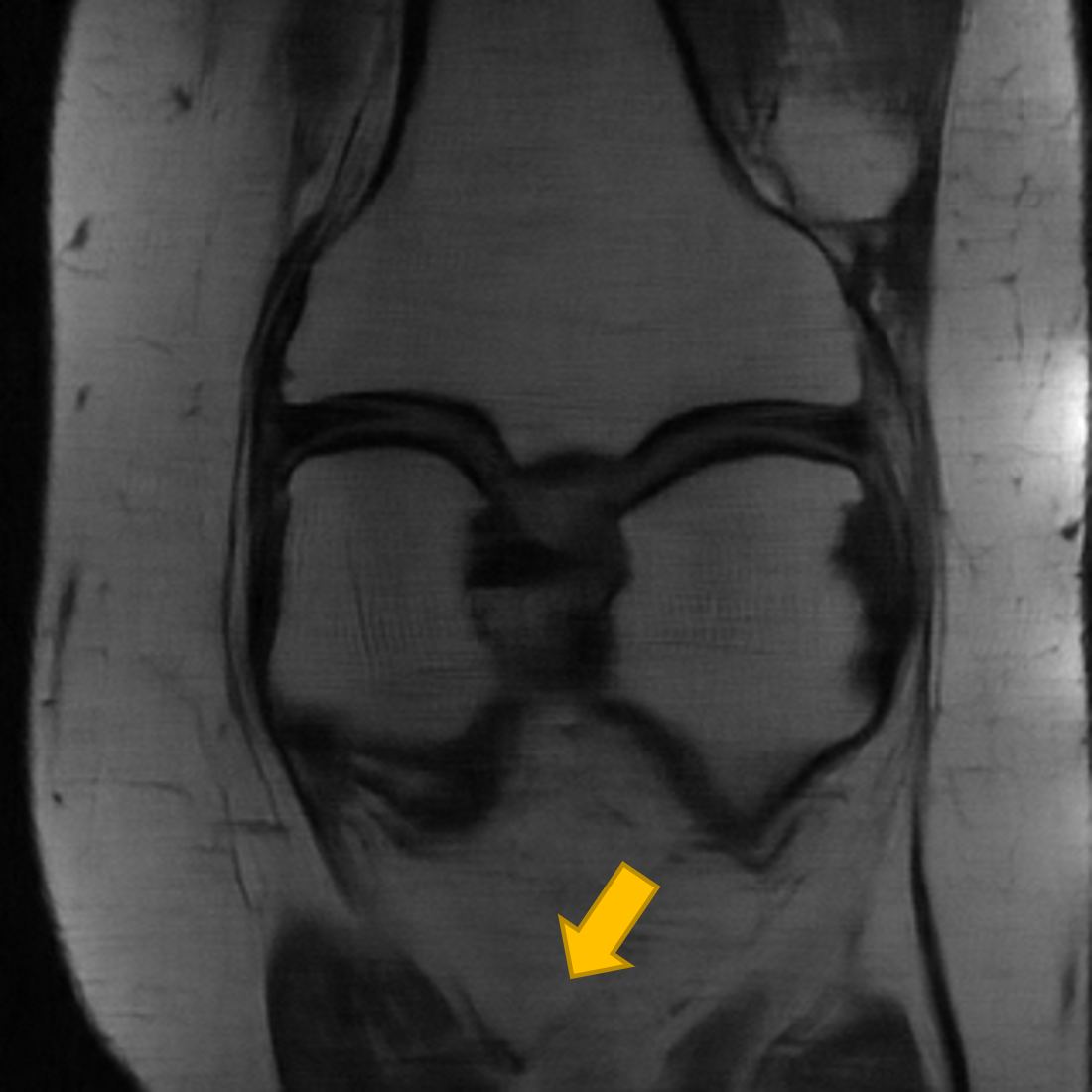}
                \caption{\scriptsize{34.3/.827}}
            \end{subfigure} &
            \begin{subfigure}[b]{0.160\linewidth}
                \includegraphics[width=\textwidth,height=\textwidth]{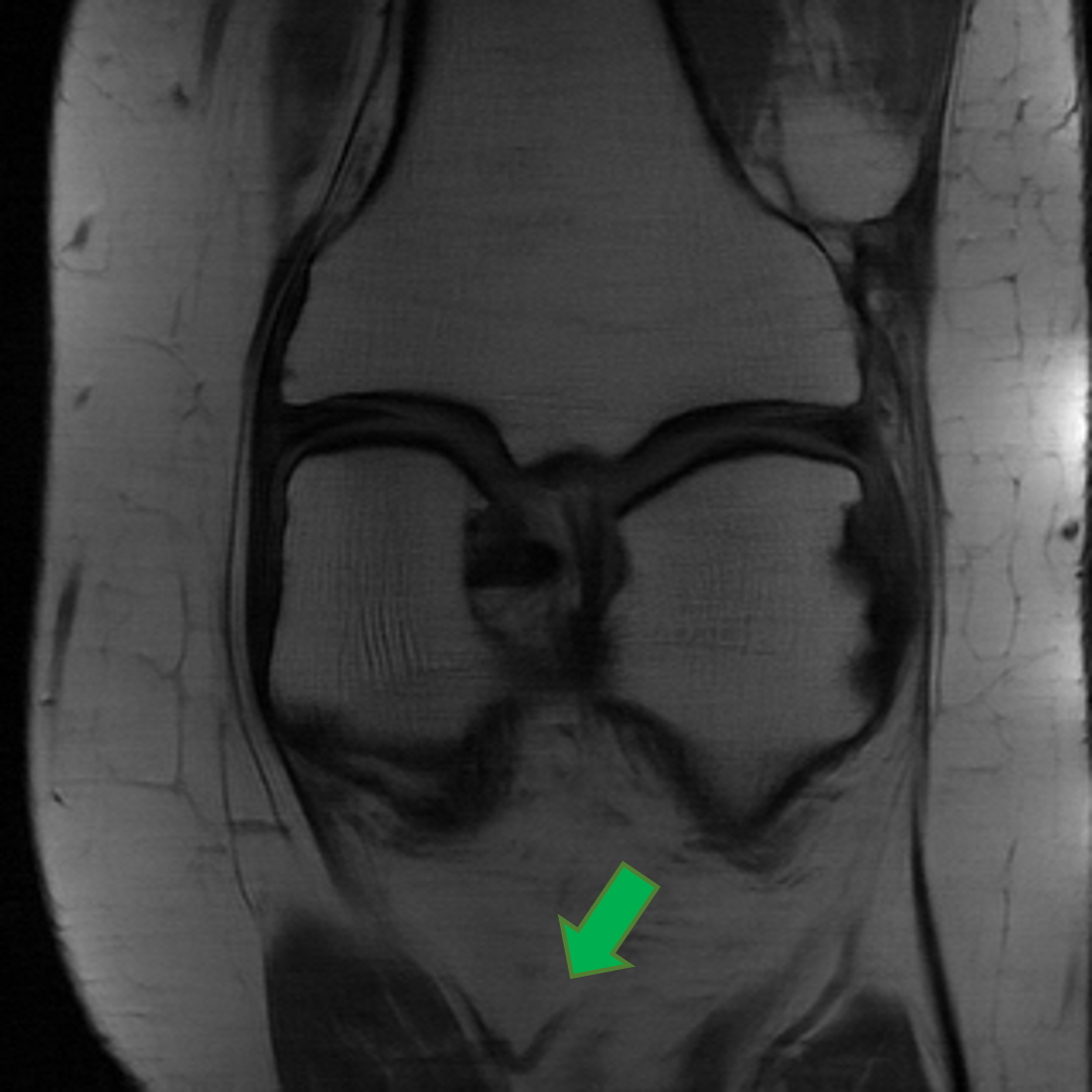}
                \caption{\scriptsize{\textbf{35.6/.864}}}
            \end{subfigure} &
            \begin{subfigure}[b]{0.160\linewidth}
                \includegraphics[width=\textwidth,height=\textwidth]{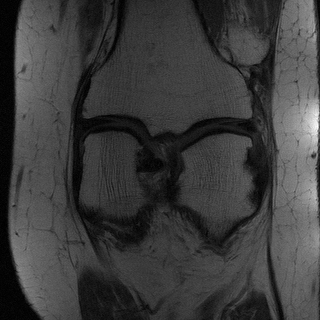}
                \caption{\scriptsize{}}
            \end{subfigure} \\
            \hline
            \multicolumn{4}{|c|}{SKM-TEA, 6$\times$$\rightarrow$10$\times$}\\
            \begin{subfigure}[b]{0.160\linewidth}
                \includegraphics[width=\textwidth,height=\textwidth]{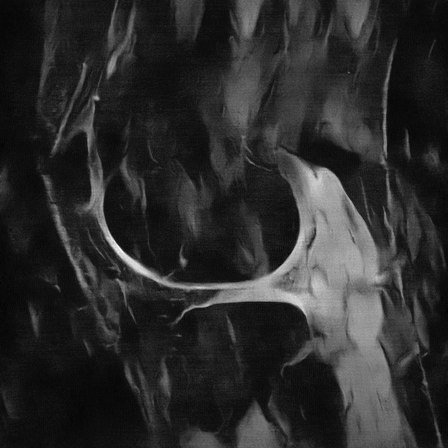}
            \caption{\scriptsize{28.1/.510}}
            \end{subfigure} &
            \begin{subfigure}[b]{0.160\linewidth}
                \includegraphics[width=\textwidth,height=\textwidth]{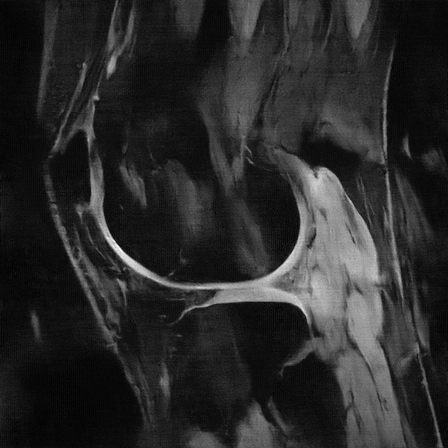}
            \caption{\scriptsize{28.2/.519}}
            \end{subfigure} &
            \begin{subfigure}[b]{0.160\linewidth}
                \includegraphics[width=\textwidth,height=\textwidth]{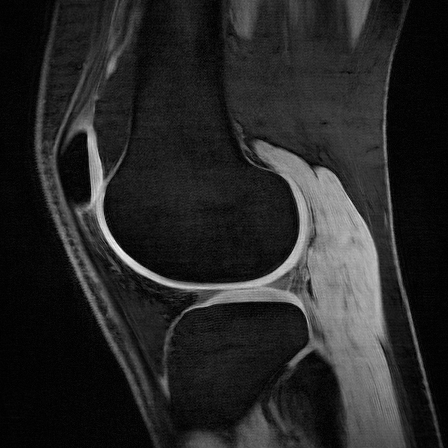}
            \caption{\scriptsize{\textbf{33.4/.736}}}
            \end{subfigure} &
            \begin{subfigure}[b]{0.160\linewidth}
                \includegraphics[width=\textwidth,height=\textwidth]{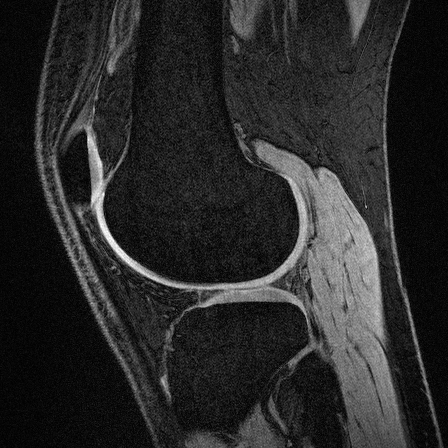}
            \caption{\scriptsize{}}
            \end{subfigure} \\  
           
            \hline        
            
        \end{tabular}

        \caption{\hspace{-9em}(b)}
        \end{subfigure}
        
    \end{tabular}     
    \caption{(a) Ablation Study on DiffuseRecon with different parameters. (b) Visualizations on recovering from unseen acceleration factors.}
    \label{fig:results}
    \vspace{-2em}
\end{figure*}

\hspace{-1.5em}\textbf{Ablation Study.} We examine the effectiveness of the following design choices:
\begin{itemize}
    \item ${\textrm{DiffuseRecon}}^{k,N}_{\textrm{nonoise}}$: $\mathcal{M}\mathcal{F}y^{\prime}_{t}$ is directly replaced with $x_{\textrm{obs}}$ in KSG instead of $x_{\textrm{obs},t}$; $k=\{40\}$ and $N=\{1,2,5,10\}$ are tested.
    \item ${\textrm{DiffuseRecon}}^{k,N}_{\textrm{norefine}}$: a combination of $\{k,N\}$ are tested without the refining steps; specifically, $k=\{8,40\}$ and $N=\{1,2,5,10\}$.
    \item ${\textrm{DiffuseRecon}}^{k,N}_{\textrm{refine}}$: $T_{\textrm{refine}}=20$ steps of further denoising is applied to the aggregate results from ${\textrm{DiffuseRecon}}^{k,N}_{\textrm{norefine}}$, $\{k,N\}=\{\{8,40\},10\}$.
\end{itemize}

The PSNR comparisons are visualized in Fig.~\ref{fig:results}(a) and are made based on the middle fifteen slices for all subjects in the fastMRI's evaluation set. There are several interesting takeaways. Firstly, PSNR increases significantly with larger $N$ for all instances, demonstrating the importance of multi-sample aggregation. We note that $k=8$ applies 500 steps and yields sufficiently denoised results; as such, the low PSNR value for ${\textrm{DiffuseRecon}}^{8,1}_{\textrm{norefine}}$ is due to the geometric variability between sampled and groundtruth image. Secondly, when no gradual noise is added to $x_{\textrm{obs}}$, the reconstruction results are significantly better when $N=1$ and worse when $N=10$ compared to the proposed KSG; such a direct replacement approach is used in a concurrent work POCS~\cite{chung2021scorebased}. This indicates that, while the clean $x_{\textrm{obs}}$ introduced at $t=T$ helps accelerate the denoising process, the aggregate results do not estimate the ground-truth well, i.e. they are more likely to be biased. Finally, the refinement steps significantly remove the blurriness caused by averaging multiple samples. We note that performances are very similar for ${\textrm{DiffuseRecon}}^{8,10}_{\textrm{refine}}$ and ${\textrm{DiffuseRecon}}^{40,10}_{\textrm{refine}}$; as such, $k=8$ leads to significant speed-up without sacrificing reconstruction quality.

\begin{table}[!htb]
\caption{Quantitative volume-wise evaluation of DiffuseRecon against SoTA methods. Dedicated models are trained for 6$\times$ and 8$\times$ acceleration; model robustness is tested by applying $\{6\times,8\times\}$ model on $\{10\times,4\times\}$ downsampled inputs. All baselines achieved similar performance compared to the original papers.}
\setlength{\tabcolsep}{3.5pt}
\label{tab:sota_table}
\centering 
\resizebox{\textwidth}{!}{%
\begin{tabular}{|c|cccccccc|}
\hline
\multirow{2}{*}{Method}&\multicolumn{4}{c|}{fastMRI}&\multicolumn{4}{c|}{SKM-TEA}\\
\cline{2-9}
&\scriptsize{6$\times$}&\scriptsize{8$\times$}&\scriptsize{8$\times$$\rightarrow$4$\times$}&\scriptsize{6$\times$$\rightarrow$10$\times$}&\scriptsize{6$\times$}&\scriptsize{8$\times$}&\scriptsize{8$\times$$\rightarrow$4$\times$}&\scriptsize{6$\times$$\rightarrow$10$\times$}\\
\hline
\multirow{2}{*}{UNet~\cite{DBLP:conf/miccai/RonnebergerFB15}}&29.79&28.89&30.67&21.98&31.62&30.47&32.06&24.34\\
&0.621&0.577&0.666&0.296&0.713&0.669&0.728&0.382\\
\hline
\multirow{2}{*}{KIKI-Net~\cite{Eo2018-xi}}&29.51&28.09&30.18&22.31&31.67&30.14&32.20&24.67\\
&0.607&0.546&0.650&0.313&0.711&0.655&0.732&0.422\\
\hline
\multirow{2}{*}{D5C5~\cite{DBLP:journals/tmi/SchlemperCHPR18}}&29.88&28.63&30.90&23.07&32.22&30.86&32.99&25.99\\
&0.622&0.565&0.675&0.349&0.732&0.683&0.763&0.512\\
\hline
\multirow{2}{*}{OUCR~\cite{DBLP:conf/miccai/GuoVWZJP21}}&30.44&29.56&31.33&23.41&32.52&31.27&33.11&26.17\\
&0.644&0.600&0.689&0.371&0.742&0.696&0.766&0.516\\
\hline
\multirow{2}{*}{DiffuseRecon}&\textbf{30.56}&\textbf{29.94}&\textbf{31.70}&\textbf{27.23}&\textbf{32.58}&\textbf{31.56}&\textbf{33.76}&\textbf{28.40}\\
&\textbf{0.648}&\textbf{0.614}&\textbf{0.708}&\textbf{0.515}&\textbf{0.743}&\textbf{0.706}&\textbf{0.795}&\textbf{0.584}\\
\hline
\end{tabular}}
\vspace{-2em}
\end{table}
\hspace{-1.5em}\textbf{Quantitative and Visual Evaluation.}
We compare reconstruction results from DiffuseRecon with current SoTA methods, and summarize the quantitative results in Table~\ref{tab:sota_table}. KIKI-Net~\cite{Eo2018-xi} applies convolutions on both the image and k-space data for reconstruction. D5C5~\cite{DBLP:journals/tmi/SchlemperCHPR18} is a seminal DL-based MR reconstruction work and proposes to incorporate data consistency layers into a cascade of CNNs. OUCR~\cite{DBLP:conf/miccai/GuoVWZJP21} builds on~\cite{DBLP:journals/tmi/SchlemperCHPR18} and uses a recurrent over-complete CNN~\cite{DBLP:conf/miccai/ValanarasuSHP20} architecture to better recovery fine details.   We note that these methods use supervised learning and train dedicated models based on a fixed acceleration factor. In Table~\ref{tab:sota_table}, DiffuseRecon is compared with these specifically trained models for acceleration factor of $6\times$ and $8\times$ in k-space. Although the model for DiffuseRecon is trained for \emph{a denoising task} and has \emph{not} observed any down-sampled MR images, DiffuseRecon obtains top performances compared to the dedicated models from current SoTAs; the performance gap is larger as the acceleration factor becomes higher. To examine the robustness of dedicated models on lower and higher quality inputs, we apply models trained on $6\times$ and $8\times$ acceleration on $10\times$ and $4\times$ undersampled inputs, respectively. In these cases, the performances of DiffuseRecon are significantly higher, demonstrating that 1. models obtained from supervised training are less reliable on images with unseen levels of degradation, and 2. DiffuseRecon is a general and performant MR reconstruction method.

\begin{figure*}[htb!]
\vspace{-2em}
    \setlength{\abovecaptionskip}{3pt}
    \setlength{\tabcolsep}{1pt}
    \centering
\resizebox{\textwidth}{!}{%
    \begin{tabular}[b]{|c|c|c|c|c|c|c|c|}
        \hline
        \multicolumn{4}{|c|}{FastMRI}&\multicolumn{4}{c|}{SKM-TEA}\\
        \hline
        D5C5~\cite{DBLP:journals/tmi/SchlemperCHPR18} & OUCR~\cite{DBLP:conf/miccai/GuoVWZJP21} & DiffuseRecon & GT & D5C5~\cite{DBLP:journals/tmi/SchlemperCHPR18} & OUCR~\cite{DBLP:conf/miccai/GuoVWZJP21} & DiffuseRecon & GT\\
        
        \hline
        \multicolumn{8}{|c|}{6$\times$}\\

        \begin{subfigure}[b]{0.160\linewidth}
            \includegraphics[width=\textwidth,height=\textwidth]{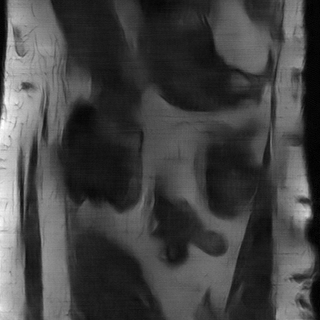}
        \end{subfigure} &
        \begin{subfigure}[b]{0.160\linewidth}
            \includegraphics[width=\textwidth,height=\textwidth]{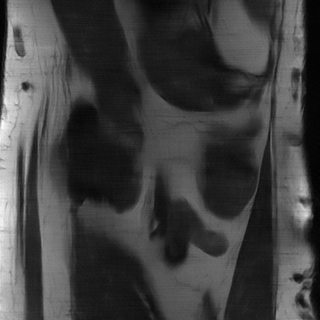}
        \end{subfigure} &
        \begin{subfigure}[b]{0.160\linewidth}
            \includegraphics[width=\textwidth,height=\textwidth]{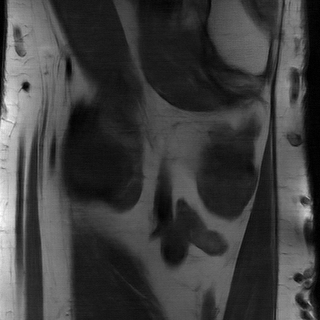}
        \end{subfigure} &
        \begin{subfigure}[b]{0.160\linewidth}
            \includegraphics[width=\textwidth,height=\textwidth]{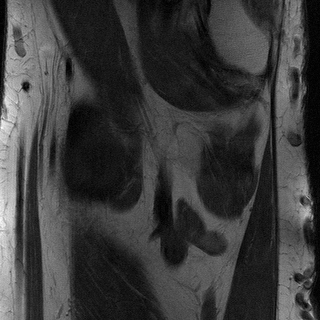}
        \end{subfigure} &
        \begin{subfigure}[b]{0.160\linewidth}
            \includegraphics[width=\textwidth,height=\textwidth]{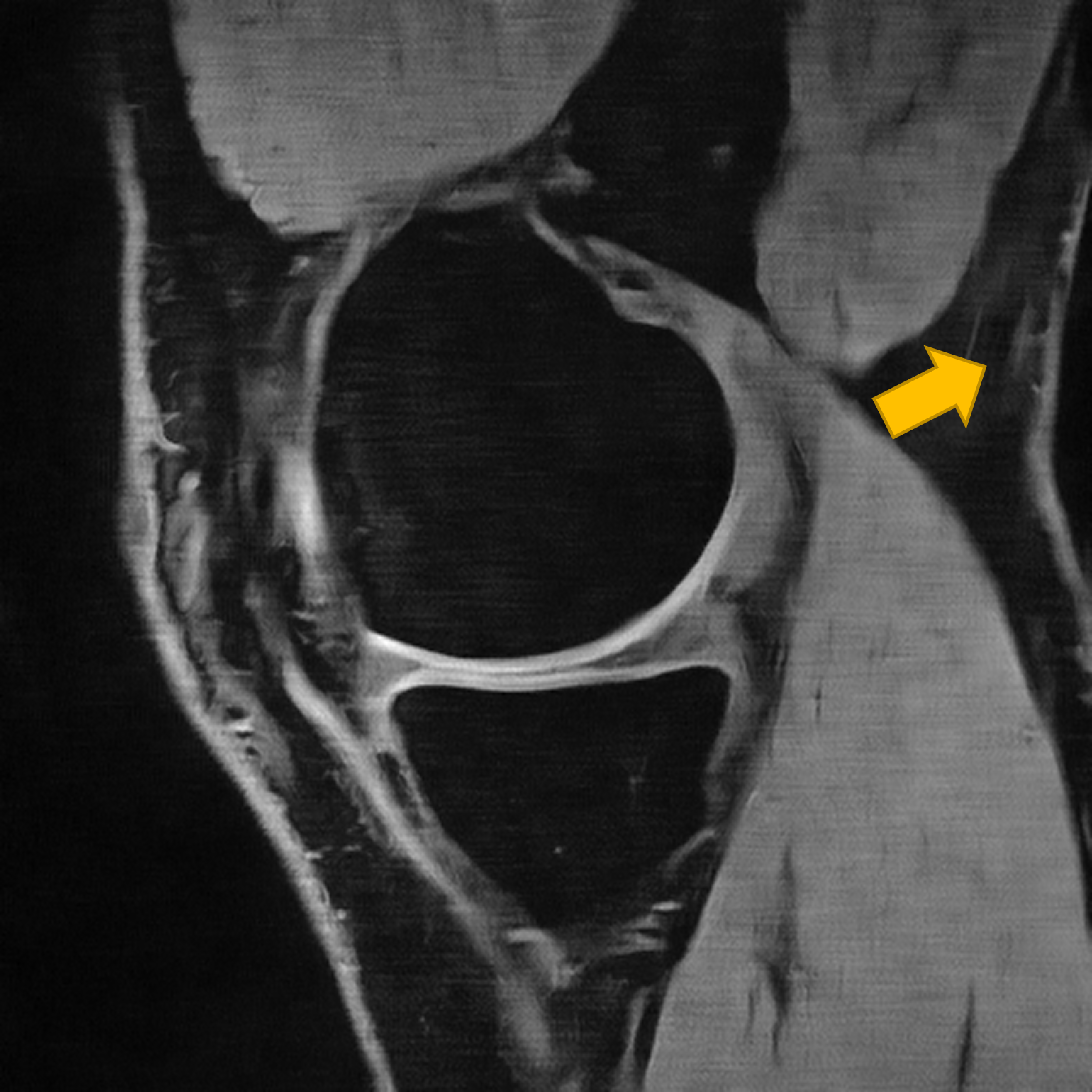}
        \end{subfigure} &
        \begin{subfigure}[b]{0.160\linewidth}
            \includegraphics[width=\textwidth,height=\textwidth]{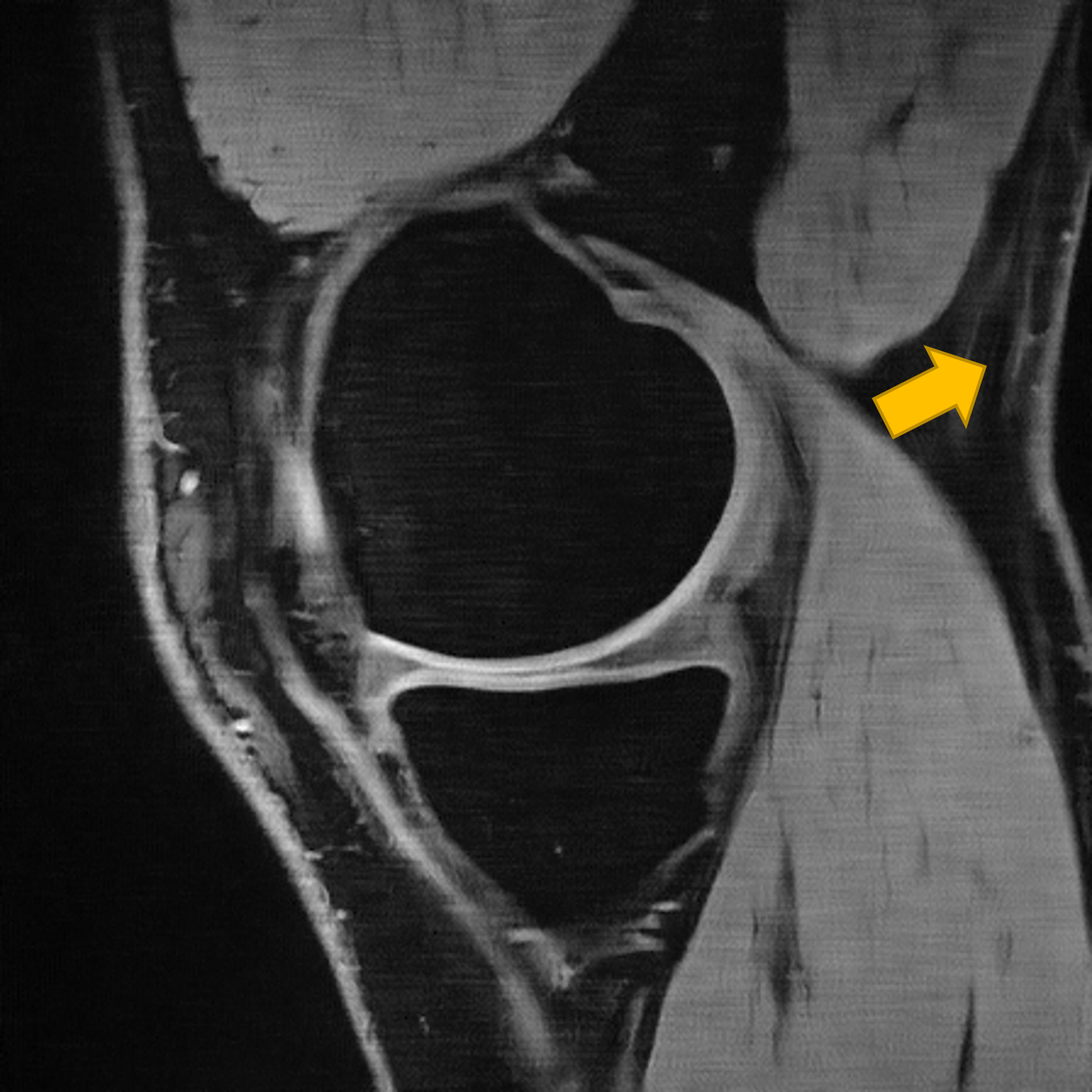}
        \end{subfigure} &
        \begin{subfigure}[b]{0.160\linewidth}
            \includegraphics[width=\textwidth,height=\textwidth]{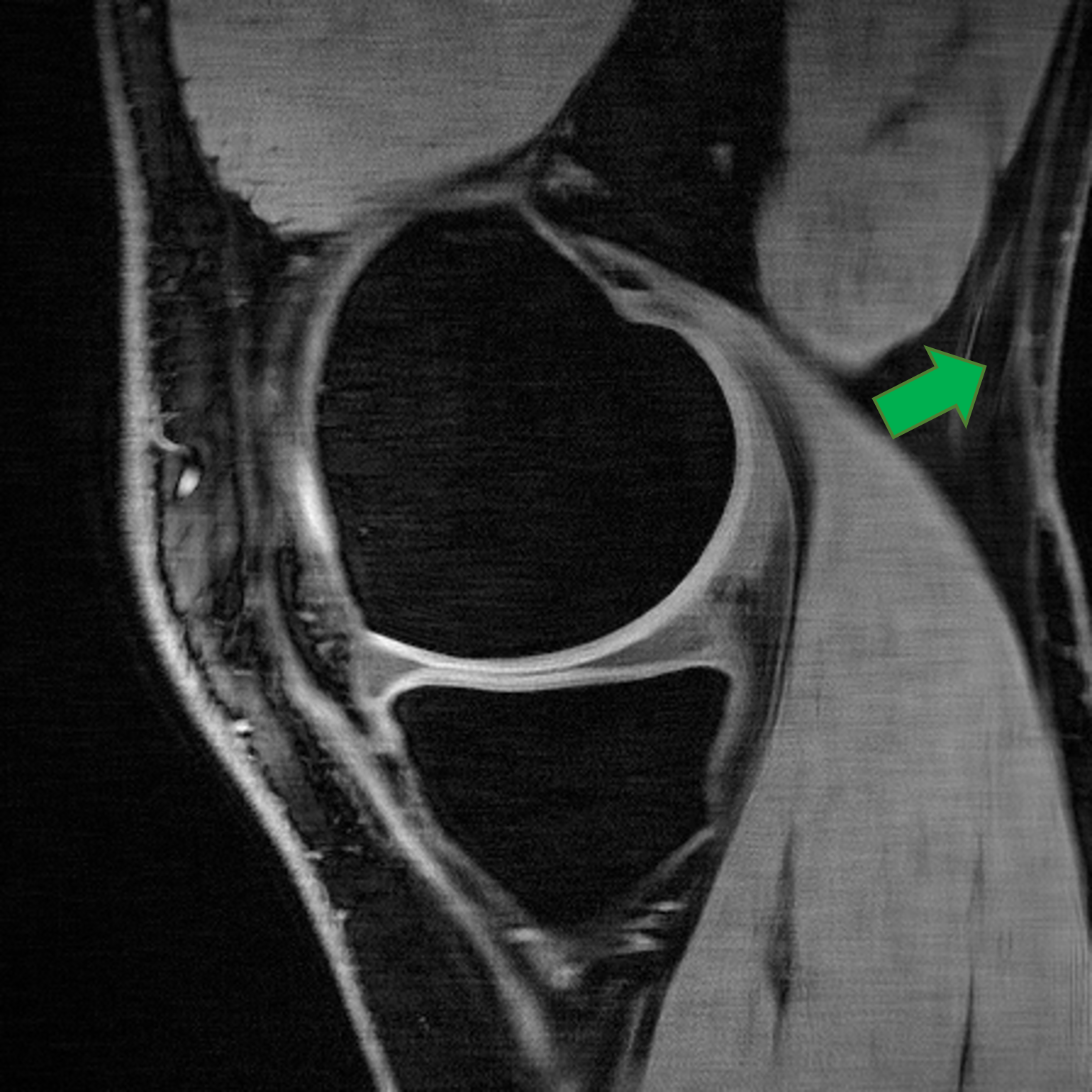}
        \end{subfigure} &
        \begin{subfigure}[b]{0.160\linewidth}
            \includegraphics[width=\textwidth,height=\textwidth]{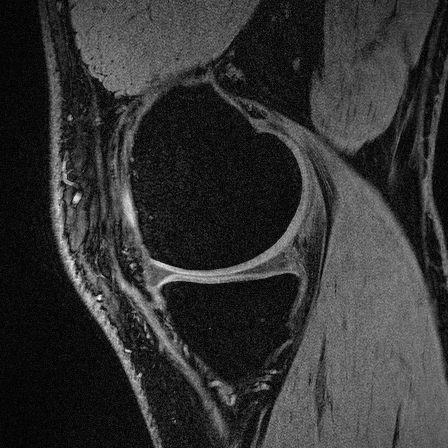}
        \end{subfigure} \\

        \begin{subfigure}[b]{0.160\linewidth}
            \includegraphics[width=\textwidth,height=\textwidth]{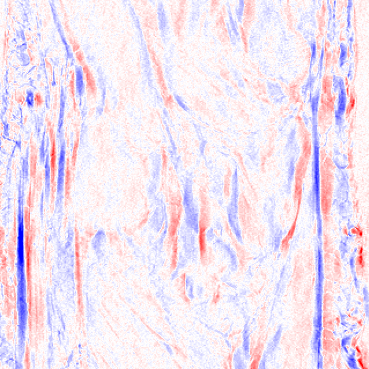}
            \caption{29.61/.6974}
        \end{subfigure} &
        \begin{subfigure}[b]{0.160\linewidth}
            \includegraphics[width=\textwidth,height=\textwidth]{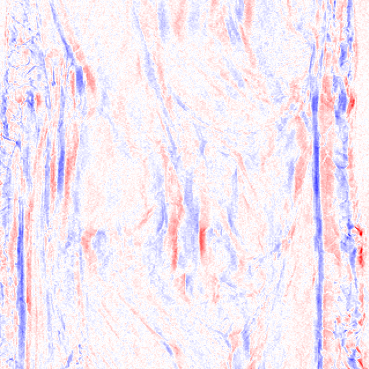}
            \caption{31.34/.7472}
        \end{subfigure} &
        \begin{subfigure}[b]{0.160\linewidth}
            \includegraphics[width=\textwidth,height=\textwidth]{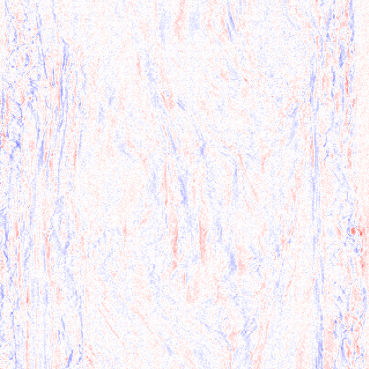}
            \caption{\textbf{35.30/.8622}}
        \end{subfigure} &
        \begin{subfigure}[b]{0.160\linewidth}
            \includegraphics[width=\textwidth,height=\textwidth]{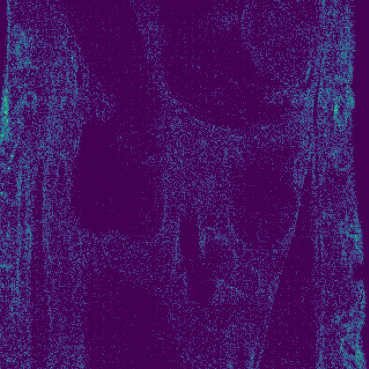}
            \caption{Variance}
        \end{subfigure} &
        \begin{subfigure}[b]{0.160\linewidth}
            \includegraphics[width=\textwidth,height=\textwidth]{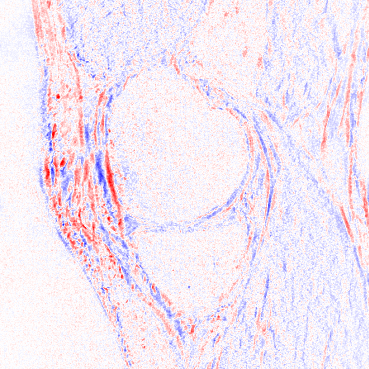}
            \caption{30.03/.6160}
        \end{subfigure} &
        \begin{subfigure}[b]{0.160\linewidth}
            \includegraphics[width=\textwidth,height=\textwidth]{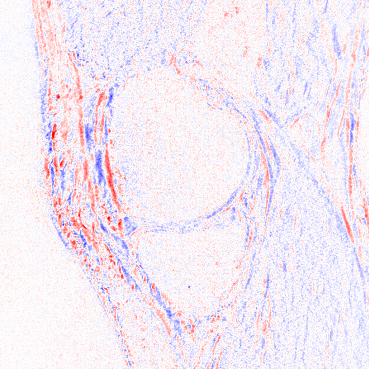}
            \caption{30.50/.6765}
        \end{subfigure} &
        \begin{subfigure}[b]{0.160\linewidth}
            \includegraphics[width=\textwidth,height=\textwidth]{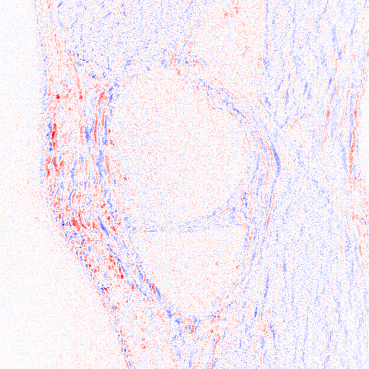}
            \caption{\textbf{31.28/.7011}}
        \end{subfigure} &
        \begin{subfigure}[b]{0.160\linewidth}
            \includegraphics[width=\textwidth,height=\textwidth]{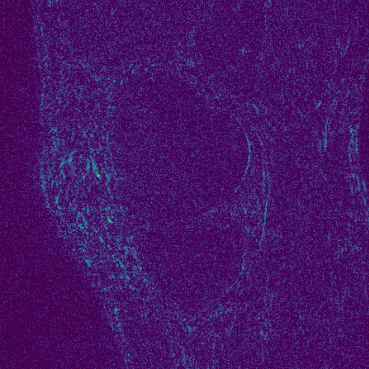}
            \caption{Variance}
        \end{subfigure}\\      

        \hline
        \multicolumn{8}{|c|}{8$\times$}\\
        \begin{subfigure}[b]{0.160\linewidth}
            \includegraphics[width=\textwidth,height=\textwidth]{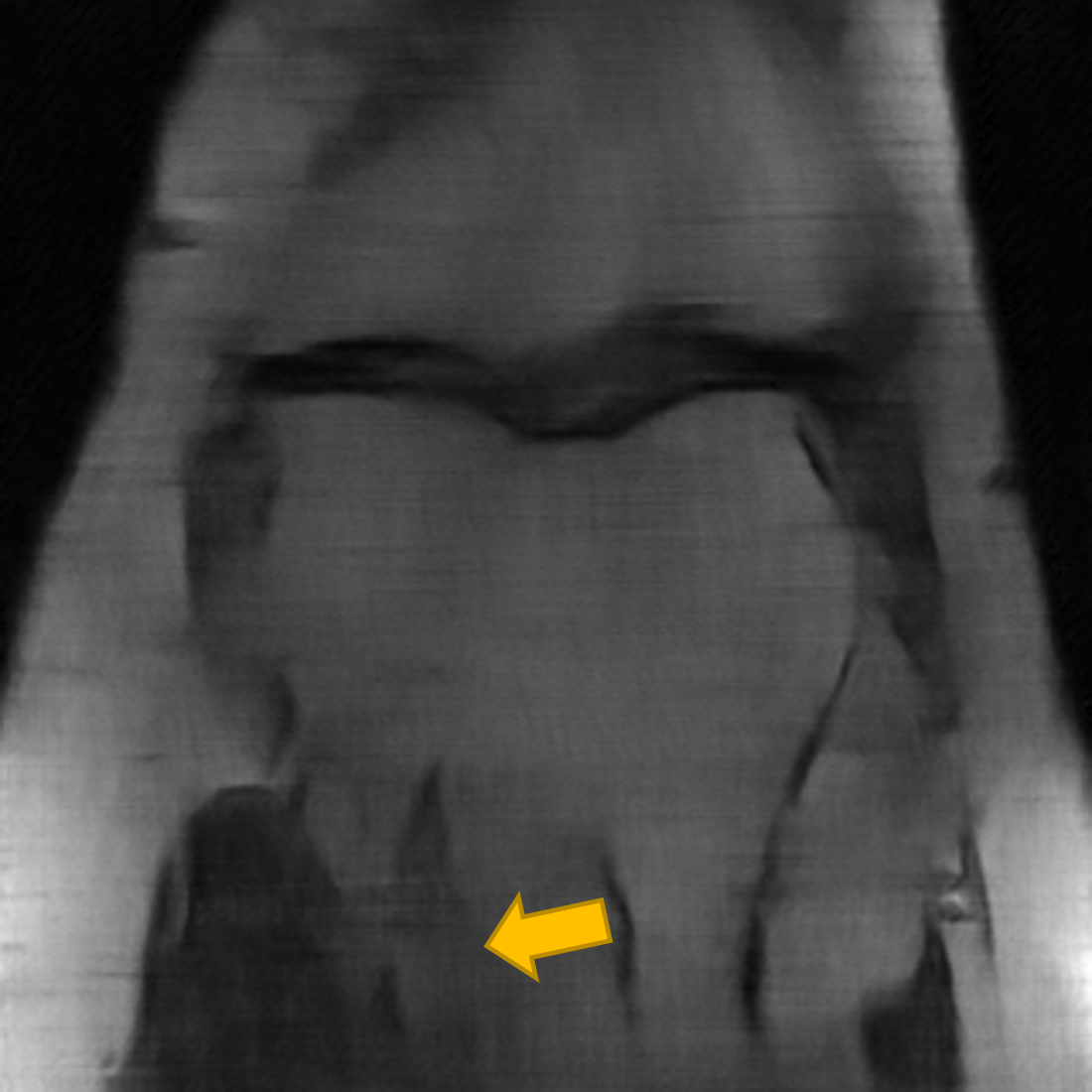}
        \end{subfigure} &
        \begin{subfigure}[b]{0.160\linewidth}
            \includegraphics[width=\textwidth,height=\textwidth]{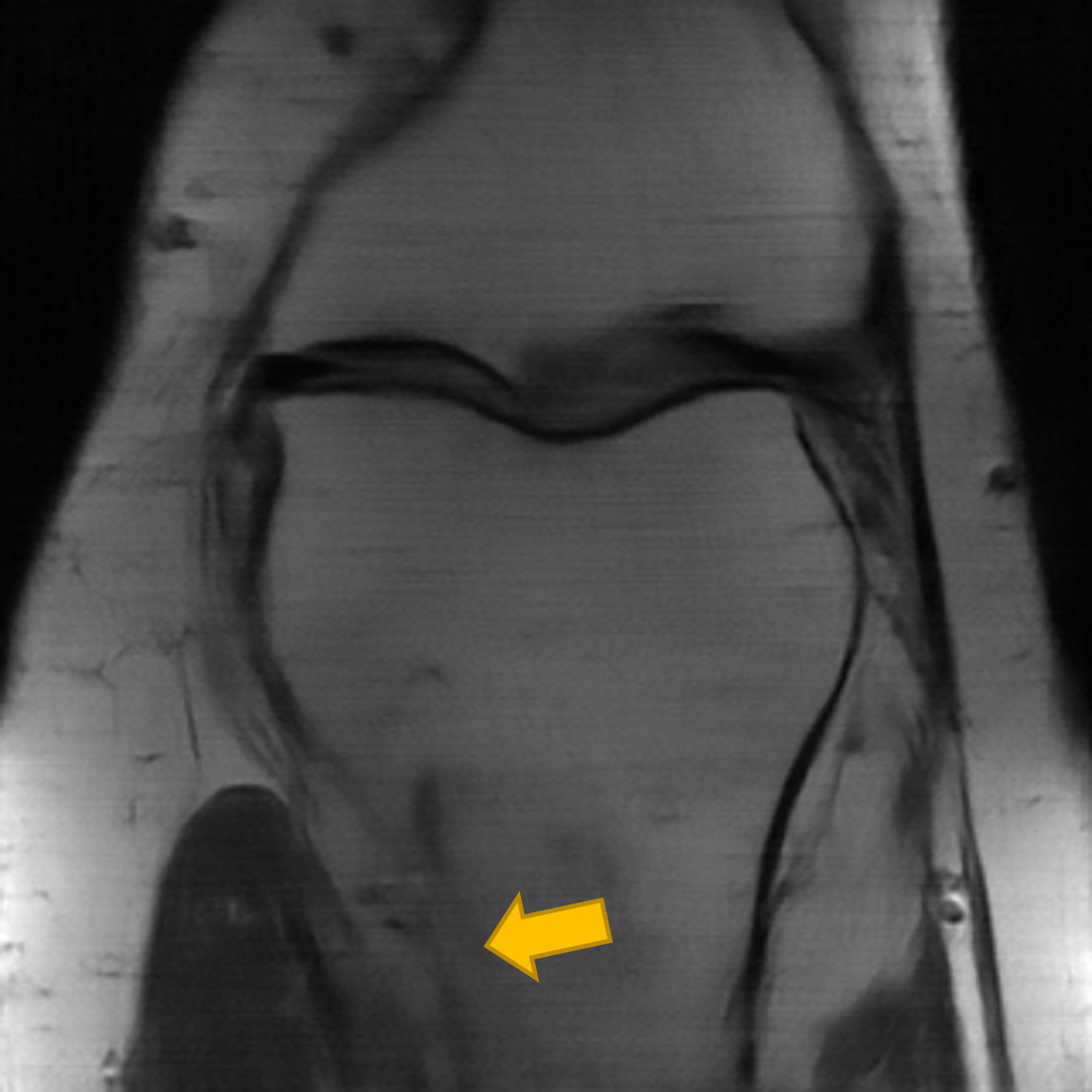}
        \end{subfigure} &
        \begin{subfigure}[b]{0.160\linewidth}
            \includegraphics[width=\textwidth,height=\textwidth]{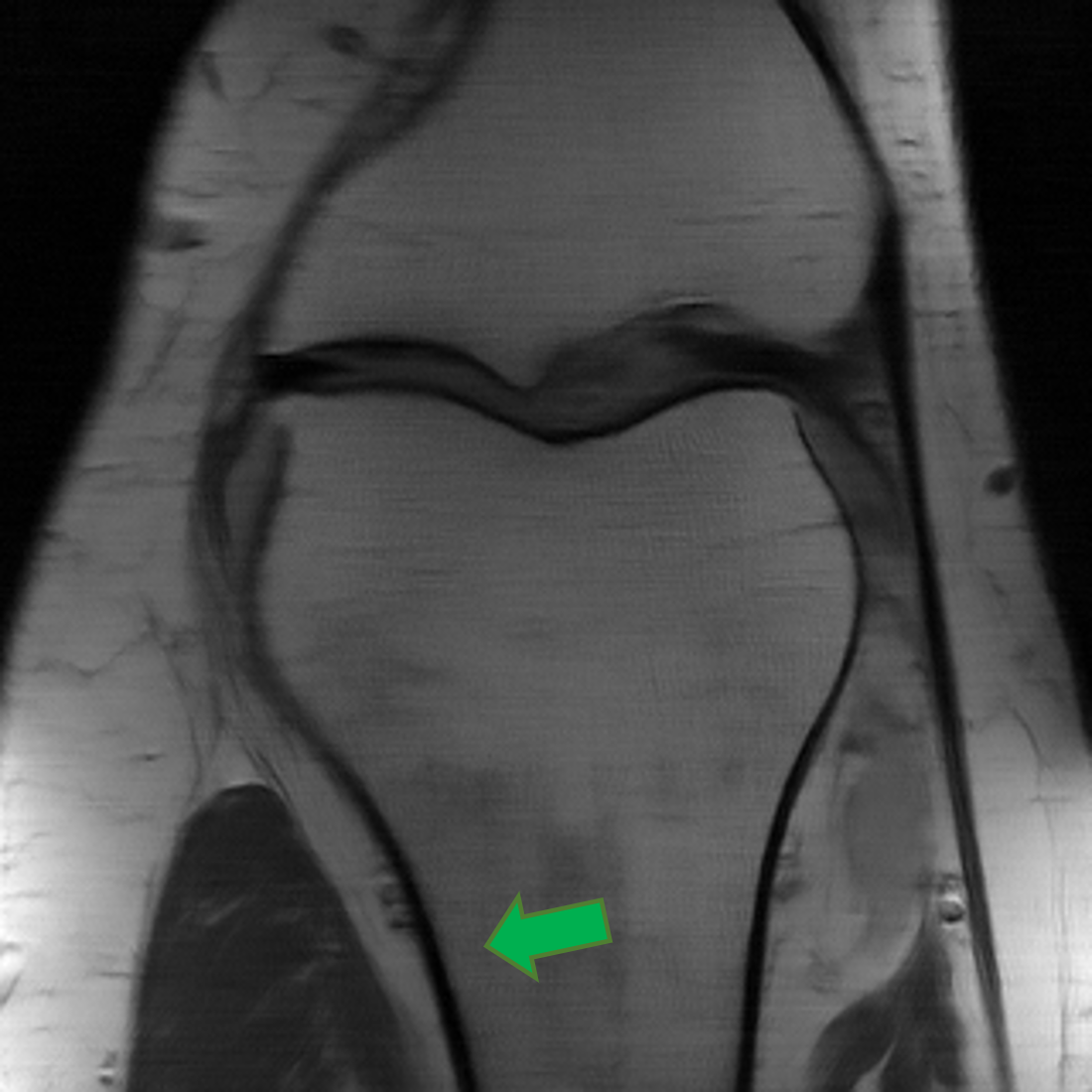}
        \end{subfigure} &
        \begin{subfigure}[b]{0.160\linewidth}
            \includegraphics[width=\textwidth,height=\textwidth]{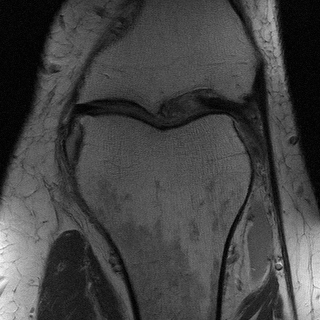}
        \end{subfigure} &
        \begin{subfigure}[b]{0.160\linewidth}
            \includegraphics[width=\textwidth,height=\textwidth]{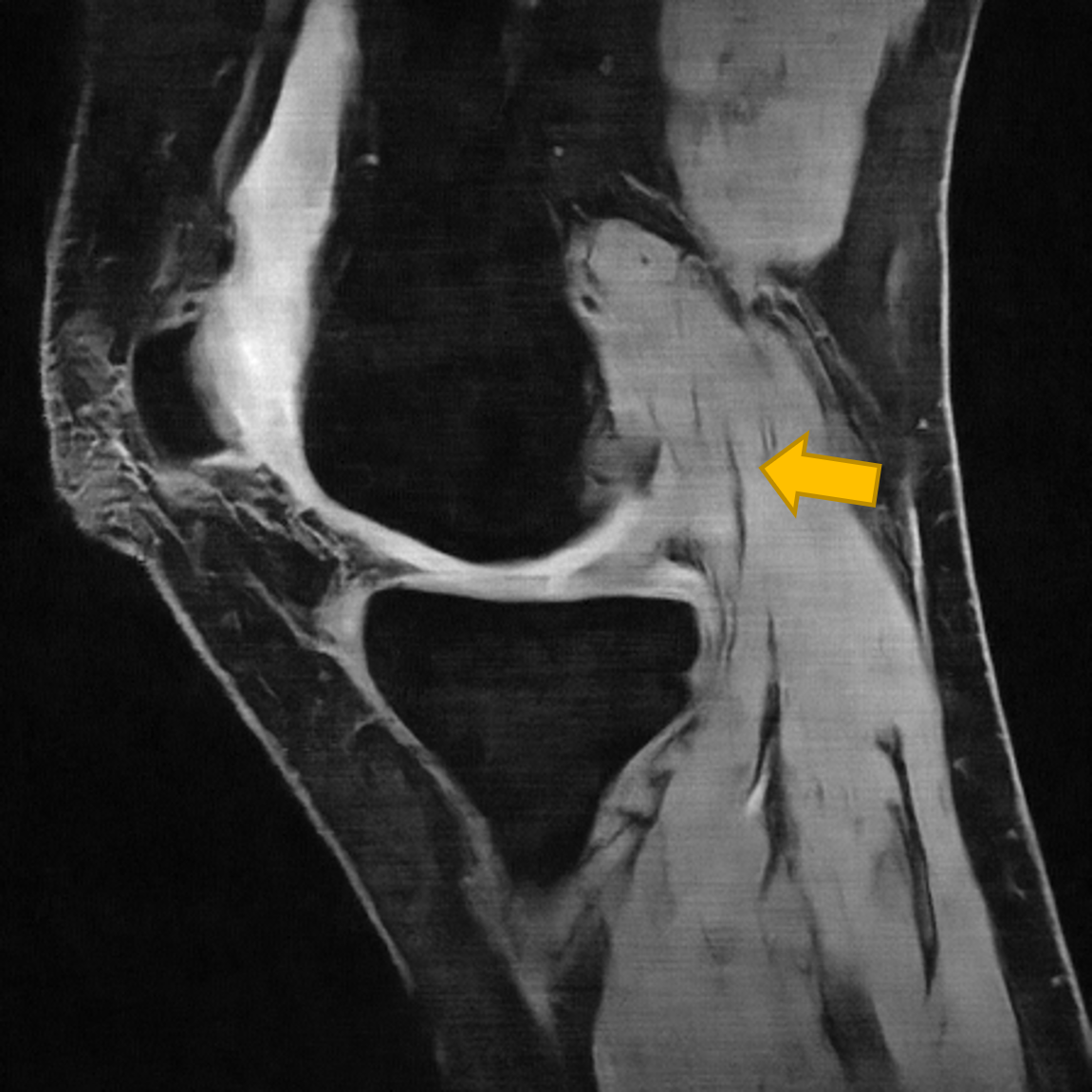}
        \end{subfigure} &
        \begin{subfigure}[b]{0.160\linewidth}
            \includegraphics[width=\textwidth,height=\textwidth]{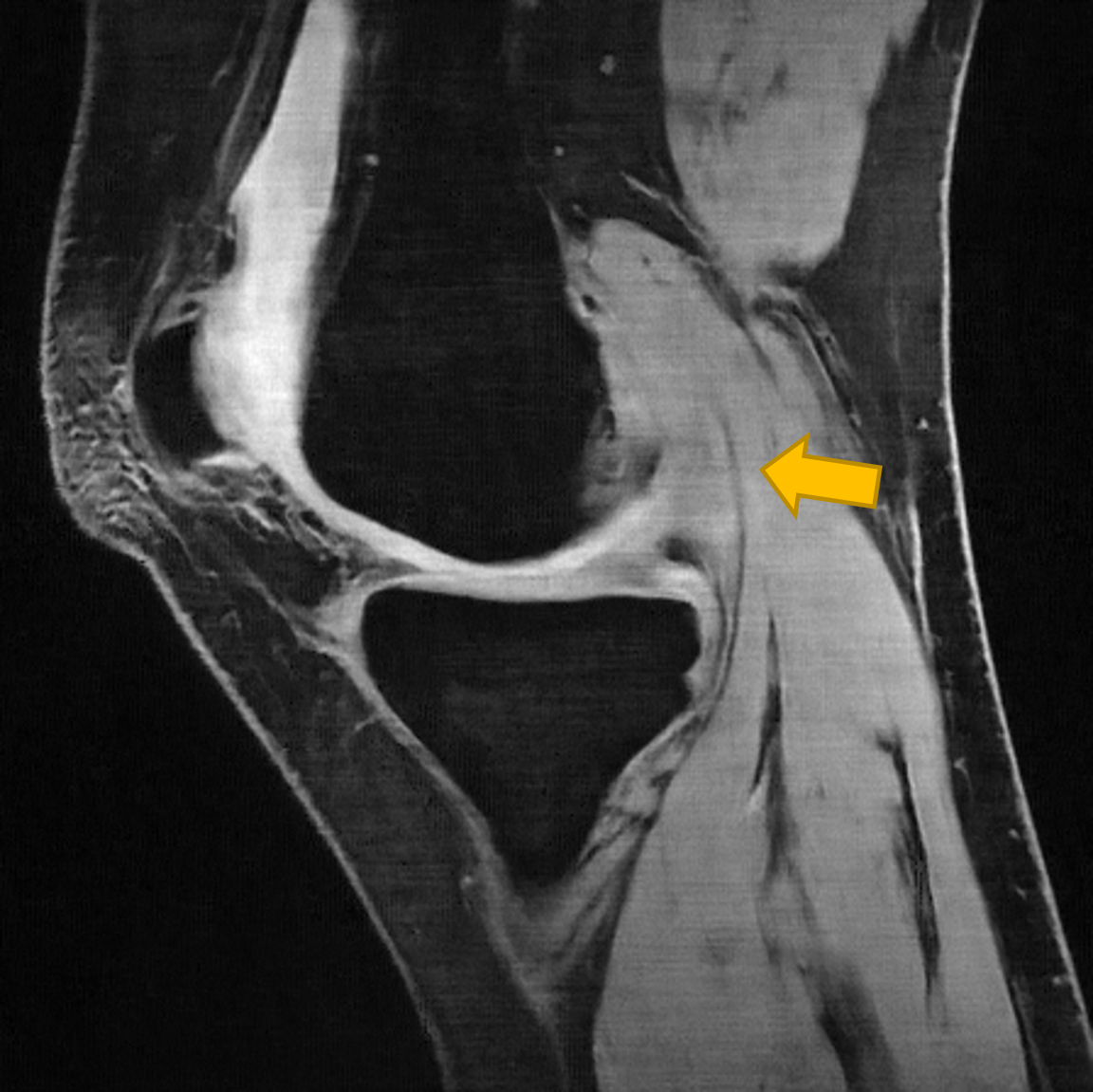}
        \end{subfigure} &
        \begin{subfigure}[b]{0.160\linewidth}
            \includegraphics[width=\textwidth,height=\textwidth]{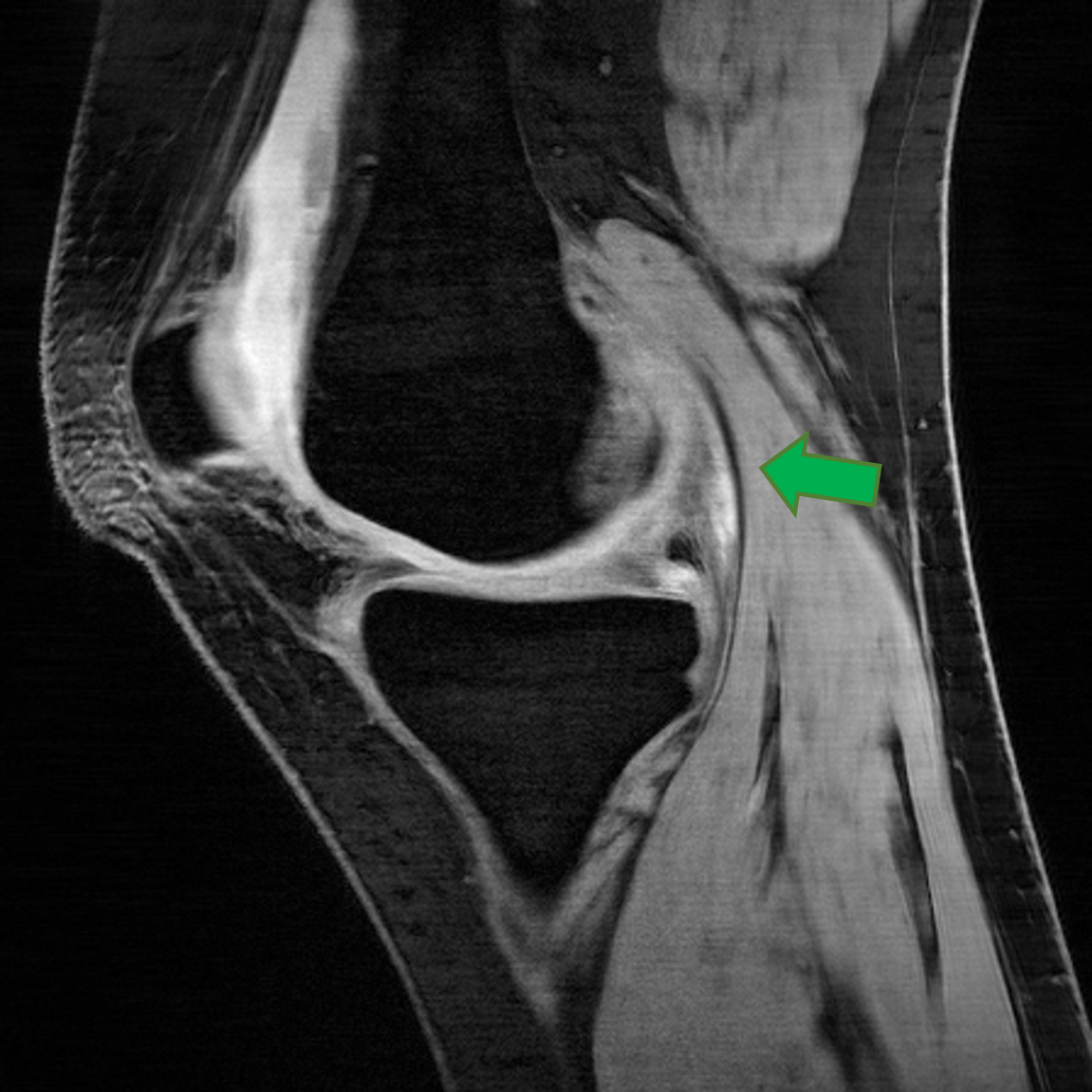}
        \end{subfigure} &
        \begin{subfigure}[b]{0.160\linewidth}
            \includegraphics[width=\textwidth,height=\textwidth]{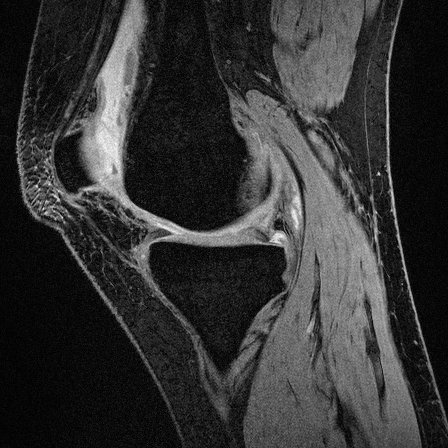}
        \end{subfigure} \\

        \begin{subfigure}[b]{0.160\linewidth}
            \includegraphics[width=\textwidth,height=\textwidth]{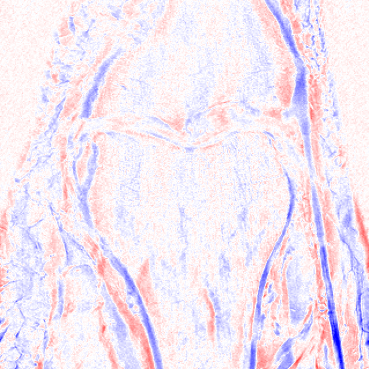}
            \caption{25.61/.5167}
        \end{subfigure} &
        \begin{subfigure}[b]{0.160\linewidth}
            \includegraphics[width=\textwidth,height=\textwidth]{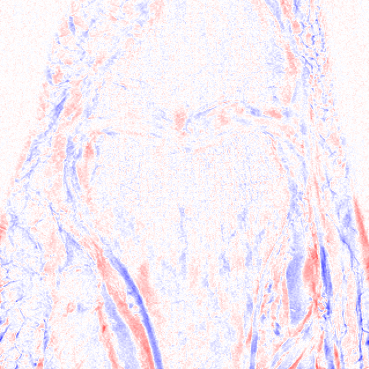}
            \caption{27.84/.6148}
        \end{subfigure} &
        \begin{subfigure}[b]{0.160\linewidth}
            \includegraphics[width=\textwidth,height=\textwidth]{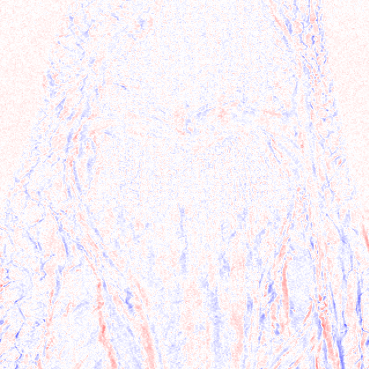}
            \caption{\textbf{30.42/.7087}}
        \end{subfigure} &
        \begin{subfigure}[b]{0.160\linewidth}
            \includegraphics[width=\textwidth,height=\textwidth]{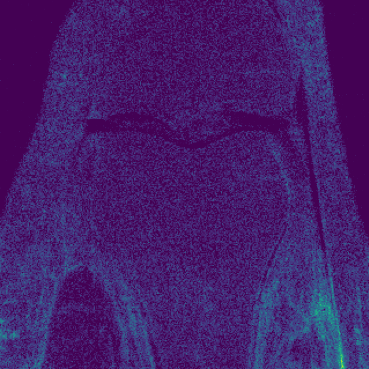}
            \caption{Variance}
        \end{subfigure} &
        \begin{subfigure}[b]{0.160\linewidth}
            \includegraphics[width=\textwidth,height=\textwidth]{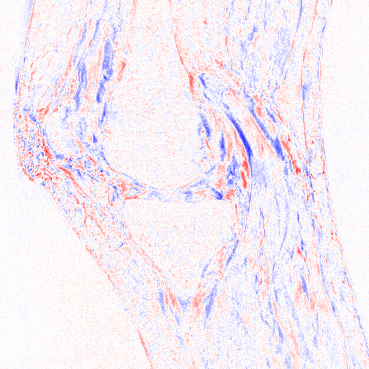}
            \caption{31.59/.7399}
        \end{subfigure} &
        \begin{subfigure}[b]{0.160\linewidth}
            \includegraphics[width=\textwidth,height=\textwidth]{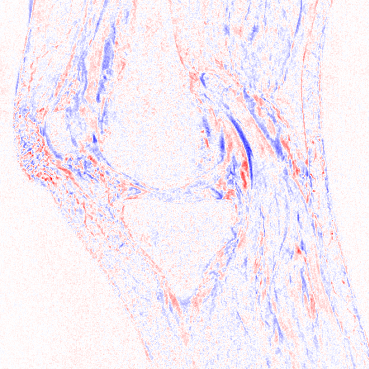}
            \caption{31.97/.7513}
        \end{subfigure} &
        \begin{subfigure}[b]{0.160\linewidth}
            \includegraphics[width=\textwidth,height=\textwidth]{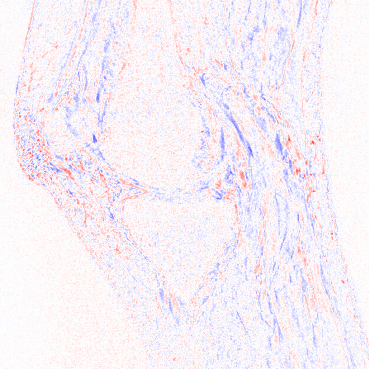}
            \caption{\textbf{33.26/.7896}}
        \end{subfigure} &
        \begin{subfigure}[b]{0.160\linewidth}
            \includegraphics[width=\textwidth,height=\textwidth]{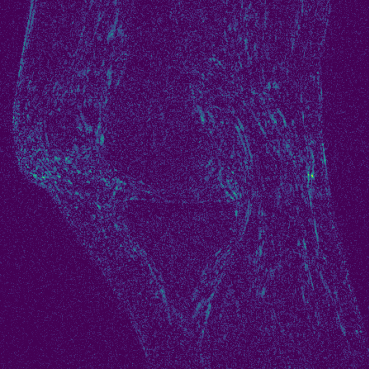}
            \caption{Variance}
        \end{subfigure}\\

        \hline        

    \end{tabular}}
    \caption{Visual comparison of DiffuseRecon with other SoTA reconstruction methods. Error maps are provided for reference.}
    \label{fig:vis_compare}
    \vspace{-3em}
\end{figure*}

Visualization of the reconstructed images is provided in Fig.~\ref{fig:results}(b) and Fig.~\ref{fig:vis_compare} to illustrate the advantages of DiffuseRecon. Due to limited space, we focus on the top three methods: D5C5~\cite{DBLP:journals/tmi/SchlemperCHPR18}, OUCR~\cite{DBLP:conf/miccai/GuoVWZJP21}, and DiffuseRecon. We observe that many significant structural details are lost in D5C5 and OUCR but can be recovered by DiffuseRecon. The errors in D5C5 and OUCR tend to have a vertical pattern, likely because the under-sampled images suffer from more vertical aliasing artifacts due to phase encoding under-sampling. As such, it is more difficult for D5C5 and OUCR to correctly recover these vertical details, and leads to blurry reconstructions under a pixel-wise loss function. DiffuseRecon, on the other hand, outputs realistic MR images that obey the distribution $p_\theta(y_{\textrm{full}}|x_{\textrm{obs}})$; it can better recover vertical details that fit the distribution of $p_\theta(y_{\textrm{full}})$. This is particularly pronounced in the $8\times$ fastMRI results in Fig.~\ref{fig:vis_compare}, where the lost vertical knee pattern in D5C5 and OUCR renders the image highly unrealistic. Such an error is avoided in DiffuseRecon, as each of its sample has a complete knee structure based on the learned prior knowledge. The uncertainty introduced by phase-encoding under-sampling is also captured by the variance map and demonstrates that the exact placement of details may be varied. For more visualization and detailed comparisons, please refer to the supplemental material.

While DiffuseRecon holds many advantages to current SoTA methods, it does require more computation due to the multi-sampling process. The one-time inference speed is comparable to UNet~\cite{DBLP:conf/miccai/RonnebergerFB15} at $20\textrm{ms}$; however, DiffuseRecon performs 1000 inference steps. We note that 20s per slice still yields significant acceleration compared to raw acquisitions, and there exists much potential in accelerating diffusion models~\cite{DBLP:conf/iclr/SongME21,DBLP:journals/corr/abs-2106-00132} for future work. DiffuseRecon can also be used in conjunction with deterministic methods, \textit{e.g.}, when variance analysis is needed only selectively, to balance speed to performance and reliability.

\captionsetup[subfigure]{labelformat=empty}
\section{Conclusion}

We propose DiffuseRecon, an MR image reconstruction method that is of high performance, robust to different acceleration factors, and allows a user to directly observe the reconstruction variations. Inspired by diffusion models, DiffuseRecon incorporates the observed k-space signals in reverse-diffusion and can stochastically generate realistic MR reconstructions. To obtain the most likely reconstruction candidate and its variance, DiffuseRecon performs a coarse-to-fine sampling scheme that significantly accelerates the generation process. We carefully evaluate DiffuseRecon on raw acquisition data, and find that DiffuseRecon achieves better performances than current SoTA methods that require dedicated data generation and training for different sampling conditions. The reconstruction variance from DiffuseRecon can help practitioners make more informed clinical decisions. Future works include further acceleration of the inference process and examination into different MR sampling patterns. 

%
%
%
%
%
\bibliographystyle{splncs04}
\bibliography{main}

\end{document}